\documentclass[utf8]{frontiersSCNS} 

\usepackage{url,hyperref,lineno,microtype,subcaption,tablefootnote}
\usepackage[onehalfspacing]{setspace}

\newcommand{\fig}[1]{{fig.\,\ref{#1}}}
\newcommand{\tab}[1]{{table\,\ref{#1}}}
\newcommand{\eqn}[1]{{eq.\,\ref{#1}}}

\newcommand{\eg}{{\it e.g.}}
\newcommand{\ie}{{\it i.e.}}
\newcommand{\etc}{{\it etc.}}

\newcommand{\aei}{(a,e,i)}

\newcommand{\deltav}{\Delta v}

\newcommand{\Lone}{L$_1$}
\newcommand{\Ltwo}{L$_2$}

\newcommand{\Lfour}{L$_4$}
\newcommand{\Lfive}{L$_5$}

\newcommand{\arcdeg}{{^{\circ}}}

\newcommand{\au}{\,\mathrm{au}}
\newcommand{\km}{\,\mathrm{km}}

\newcommand{\kps}{\,\mathrm{km}\,\mathrm{s}^{-1}}

\newcommand{\meter}{\,\mathrm{m}}
\newcommand{\cm}{\,\mathrm{cm}}

\newcommand{\yr}{\,\mathrm{yr}}
\newcommand{\Myr}{\,\mathrm{Myr}}

\newcommand{\Day}{\,\mathrm{day}}
\newcommand{\days}{\,\mathrm{d}}

\newcommand{\degrees}{\,\mathrm{deg}}
\newcommand{\degperday}{\,\mathrm{deg}\,\mathrm{day}^{-1}}
\newcommand{\hours}{\,hours}

\newcommand{\second}{\,\mathrm{s}}
\newcommand{\mps}{\,\meter\,\second^{-1}}

\newcommand{\kg}{\,\mathrm{kg}}

\newcommand{\designation}[2]{{{#1}\,{#2}}}

\newcommand{\RH}{{2006\,RH$_{120}$}}

\newcommand{\PSone}{\protect \hbox {Pan-STARRS1}}

\def\keyFont{\fontsize{8}{11}\helveticabold }
\def\firstAuthorLast{Jedicke {et~al.}} 
\def\Authors{Robert Jedicke\,$^{1,*}$, Bryce T. Bolin\,$^{2}$, William F. Bottke\,$^{3}$, Monique Chyba\,$^{4}$, Grigori Fedorets\,$^{5}$, Mikael Granvik\,$^{6,5}$, Lynne Jones\,$^{2}$, and Hodei Urrutxua\,$^{7}$}
\def\Address{$^{1}$Institute for Astronomy, University of Hawai`i, Honolulu, HI, USA \\
$^{2}$Department of Astronomy, University of Washington, Seattle, WA\\
$^{3}$Southwest Research Institute, Boulder, CO, USA \\
$^{4}$Department of Mathematics, University of Hawai`i, Honolulu, HI, USA \\
$^{5}$Department of Physics, University of Helsinki, Helsinki, Finland \\
$^{6}$Department of Computer Science, Electrical and Space Engineering, Lule\aa{} University of Technology, Kiruna, Sweden \\
$^{7}$Universidad Rey Juan Carlos, Madrid, Spain
}
\def\corrAuthor{Corresponding Author}

\def\corrEmail{jedicke@hawaii.edu}

\begin{document}
\onecolumn
\firstpage{1}

\title[Earth's Minimoons]{Earth's Minimoons:\\Opportunities for Science and Technology.} 

\author[\firstAuthorLast ]{\Authors} 
\address{} 
\correspondance{} 

\extraAuth{}

\maketitle

\begin{abstract}
\section{}
Twelve years ago the Catalina Sky Survey discovered Earth's first known natural geocentric object other than the Moon, a few-meter diameter asteroid designated \RH.  Despite significant improvements in ground-based telescope and detector technology in the past decade the asteroid surveys have not discovered another temporarily-captured orbiter (TCO; colloquially known as minimoons) but the all-sky fireball system operated in the Czech Republic as part of the European Fireball Network detected a bright natural meteor that was almost certainly in a geocentric orbit before it struck Earth's atmosphere.  Within a few years the Large Synoptic Survey Telescope (LSST) will either begin to regularly detect TCOs or force a re-analysis of the creation and dynamical evolution of small asteroids in the inner solar system.

The first studies of the provenance, properties, and dynamics of Earth's minimoons suggested that there should be a steady state population with about one 1- to 2-meter diameter captured objects at any time, with the number of captured meteoroids increasing exponentially for smaller sizes.  That model was then improved and extended to include the population of temporarily-captured flybys (TCFs), objects that fail to make an entire revolution around Earth while energetically bound to the Earth-Moon system.  Several different techniques for discovering TCOs have been considered but their small diameters, proximity, and rapid motion make them challenging targets for existing ground-based optical, meteor, and radar surveys.  However, the LSST's tremendous light gathering power and short exposure times could allow it to detect and discover many minimoons.

We expect that if the TCO population is confirmed, and new objects are frequently discovered, they can provide new opportunities for 1) studying the dynamics of the Earth-Moon system, 2) testing models of the production and dynamical evolution of small asteroids from the asteroid belt, 3) rapid and frequent low delta-v missions to multiple minimoons, and 4) evaluating in-situ resource utilization techniques on asteroidal material.

Here we review the past decade of minimoon studies in preparation for capitalizing on the scientific and commercial opportunities of TCOs in the first decade of LSST operations. 

\tiny
 \keyFont{ \section{Keywords:} minimoon, asteroid, NEO, ISRU, dynamics} 
\end{abstract}

\section{Minimoon introduction}
\label{s.MinimoonIntroduction}

For more than four billion years the Earth has been accompanied by the $\sim3,500\km$ diameter Moon, its only permanent natural satellite.  Our outsized satellite places the Earth at the top of the list of the eight planets in the Solar System in terms of the primary-to-satellite mass ratio despite the fact that the Moon is only about 1\% of Earth's mass.  This work reviews the history, properties, and future potential of natural objects that are {\it temporarily} gravitationally bound within the Earth-Moon system (EMS).  We refer to them as either temporarily captured objects (TCO) or temporarily captured flybys (TCF) depending on whether they make at least one revolution around Earth (the definition will be refined in \S\ref{s.MinimoonDynamics}).  As an homage to the Moon and Austin Powers\footnote{A fictional secret agent played by the Canadian comedian Mike Myers.}  we usually refer to TCOs and TCFs as `minimoons' though, to be more precise based on their relative diameters, they may more accurately be considered micromoons.

\begin{figure}[htbp]
\begin{center}

    \includegraphics[width=0.9\columnwidth]{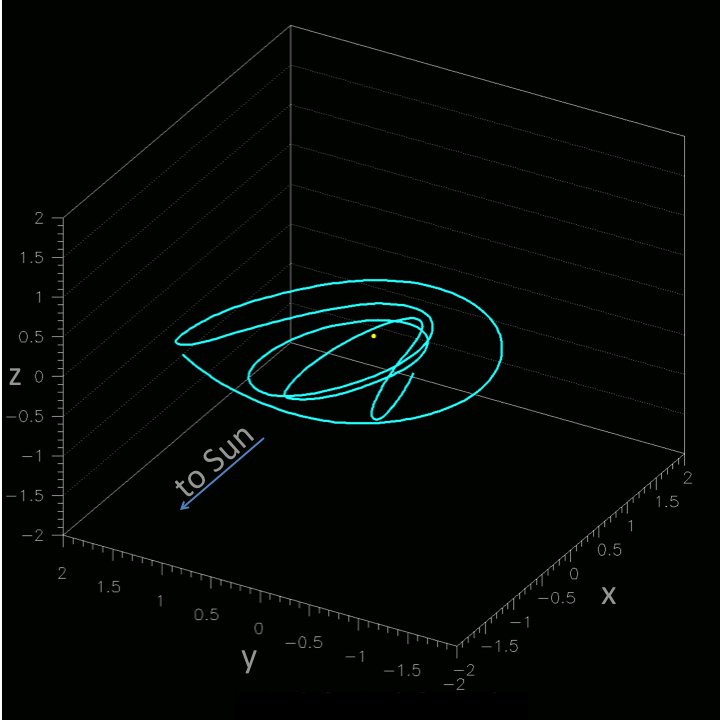}
    \includegraphics[width=0.9\columnwidth]{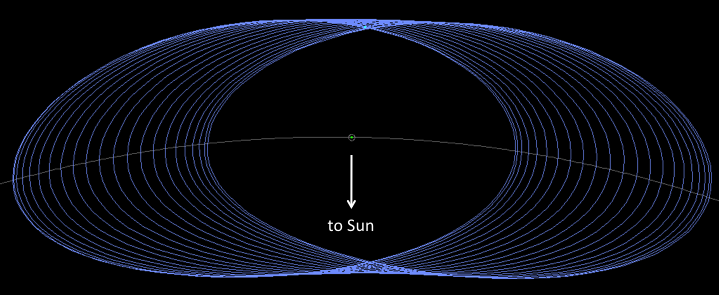}
    
    \caption{Minimoons (temporarily captured objects, TCOs) are gravitationally bound to the Earth-Moon system while quasi-satellites (\S\ref{s.MinimoonSourcePopulation}) are not.  ({\bf top}) Trajectory of the minimoon \designation{2006}{RH$_{120}$} during its capture in the Earth-Moon system in 2006-2007.  The Earth is represented by the yellow dot located at the origin of the J2000.0 mean equator and equinox reference system.  ({\bf bottom})  Trajectory of Earth's quasi-satellite \designation{2016}{HO$_3$} shown in blue as projected onto the heliocentric ecliptic $x-y$ plane in the synodic frame.  Earth is represented by the green dot in the centre and the Moon's orbit is represented by the small white circle.  Earth's orbit is shown as as the white arc from left to right and the direction to the Sun is to the bottom. \citep[credit: Paul Chodas (NASA/JPL);][]{Chodas2016DPS-HO3}.}
    \label{fig.Minimoon-vs-QuasiSatellite}

\end{center}
\end{figure}

The most basic definition of whether two objects are gravitationally bound to one another requires that the sum of their relative kinetic and potential energy must be less than zero.  \ie\
\begin{equation}
    \epsilon = \frac{E_T}{m} = \frac{c_3}{2} = \frac{1}{2} v^2 - \frac{\mu}{r} < 0
    \label{eqn.SpecificOrbitalEnergy}
\end{equation}
where $\epsilon$ is an object's specific orbital energy, the total energy ($E_T$) per unit mass ($m$) of the smaller object, $c_3$ is its `characteristic energy', $v$ and $r$ are the relative speed and distance between the objects, and $\mu = GM$ is the standard gravitational parameter where $G$ is the gravitational constant and $M$ is the mass of the primary.  This definition breaks down when there are more than two objects (\ie\ in all real situations) and in our Solar System `temporary capture' usually also requires a limit on the separation between the objects of less than 3 Hill radii \citep[\eg][]{Kary1996-SPC-CaptureStatistics,Granvik2012}.  Minimoons are temporarily captured natural satellites of Earth in the sense that they have $\epsilon<0$ with respect to Earth {\it and} are within 3 Hill radii (\fig{fig.Minimoon-vs-QuasiSatellite}).

The existence of minimoons was long regarded as impossible or, at best, unlikely, because several long-running asteroid surveys had not identified any natural geocentric objects in many years of operation.  We think this is most likely due to these objects typically being too small, too faint, and moving too rapidly to be efficiently detected, but there is also likely a psychological bias against their discovery that still remains.  Since it is `well known' that Earth has no other natural satellites any geocentric object must be artificial even if it was identified on an unusual distant orbit.  In this work we will show that this bias is unwarranted, minimoons have been discovered and will be discovered in even greater numbers in the near future as highly capable astronomical surveys begin their operations.

\section{Minimoon discoveries}
\label{s.MinimoonDiscoveries}

The Catalina Sky Survey \citep{Larson1998} has been in operation for about 20 years and has discovered many near-Earth objects (NEO; objects with perihelia $q<1.3\au$) and comets but in September 2006 they discovered the first verified minimoon\footnote{MPEC 2008-D12; \url{https://www.minorplanetcenter.net/mpec/K08/K08D12.html}}, now known as \designation{2006}{RH$_{120}$} \citep{Kwiatkowski2009}.  While its geocentric orbit was established soon after discovery there was some controversy over its nature as an artificial or natural object.  Several launch vehicle booster stages have achieved sufficient speed for them to escape the gravitational bonds of the EMS \citep[\eg][]{Jorgensen2003-J002E3} only to be subsequently recaptured in the system after a few decades.  Subsequent astrometric observations of \designation{2006}{RH$_{120}$} established its provenance as a natural object because the perturbations to its trajectory caused by solar radiation  pressure\footnote{\url{https://echo.jpl.nasa.gov/asteroids/6R10DB9/6R10DB9_planning.html}} were inconsistent with it being artificial \citep{Kwiatkowski2009}.  Later radar observations established that it is a few meters in diameter \citep{Benner2015-AsteroidsIV}.   \designation{2006}{RH$_{120}$} remained bound in the Earth-Moon system for about a year during which it made about four revolutions around the geocenter (\fig{fig.Minimoon-vs-QuasiSatellite}).  Its pre-capture orbit had a semi-major axis of $(a,e,i)\sim(0.95\au,0.05,0.6\arcdeg)$ so its aphelion was near Earth's orbit while its post-capture orbit has a perihelion close to $1\au$ with $(a,e,i)\sim(1.03\au,0.03,0.6\arcdeg)$ \citep{Granvik2012}.  We will show below that \designation{2006}{RH$_{120}$}'s dynamical properties make it a poster child for minimoon behaviour while asteroids close to its few-meter diameter should be captured with decadal frequency.

\begin{figure}[htbp]
\begin{center}
    \includegraphics[width=\columnwidth]{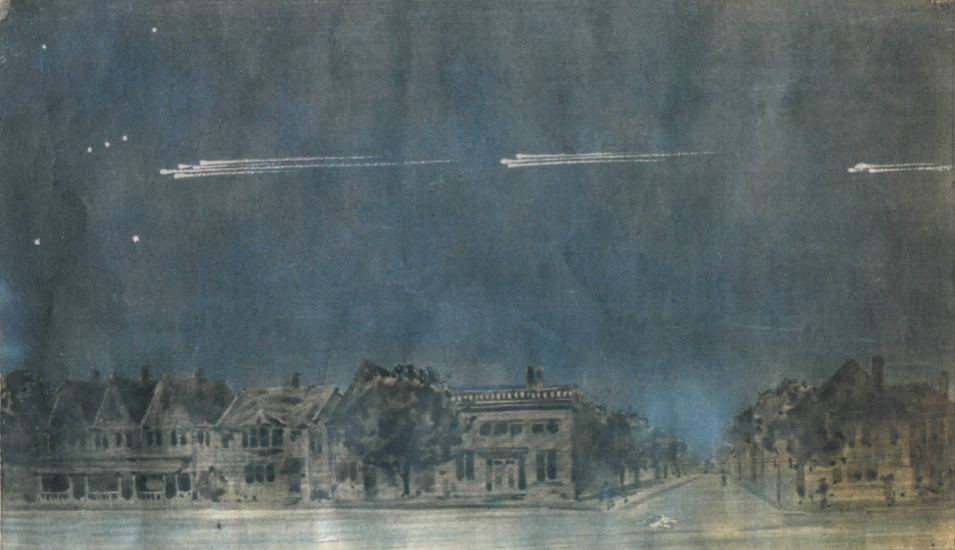}
    \caption{On 9 February 1913 ``The [Toronto] Globe [newspaper] office was flooded with reports of `a meteoric performance of stupendous dimensions'" \citep{Semeniuk2013}.  Toronto artist Gustav Hahn witnessed the minimoon fireball procession of 1913 and later painted it.  (University of Toronto Archives (A2008-0023) Copyright Natalie McMinn.)
    The first meteor photograph was obtained in 1885 \citep{Weber2005-MeteorApparatus} but eyewitness accounts and paintings were acceptable forms of observational evidence in the early 20th century.}
    \label{fig.ChantProcession}
\end{center}
\end{figure}

While \designation{2006}{RH$_{120}$} is undoubtedly the first verified minimoon discovered while in its TCO phase there are other significant minimoon observations.  The first was ``The Extraordinary Meteoric Display'' on 9 February 1913 that was observed from Saskatchewan to Bermuda (\fig{fig.ChantProcession}) and was described and analyzed by \citet{Chant1913a,Chant1913b}.  Historical researchers have even identified sightings of the event off the coast of Brazil \citep{Olson2013}!  The meteor display included dozens, and perhaps hundreds, of fragments that moved {\it slowly} across the sky in ``perfect formation''.  They were not the typical shooting star that last for only a fraction of second --- the entire procession lasted more than three minutes!  Witnesses reported that the meteors caused a ``rumbling noise'' and houses to shake along the path.  Chant's detailed analysis of eyewitness reports concluded that the object's speed with respect to Earth's surface was between $8\kps$ and $16\kps$ while Earth's escape speed or, equivalently, the speed at which an object with zero relative speed at infinity would strike Earth, is about $11.2\kps$.  He thus concluded that the meteoroid ‘‘had been traveling through space, probably in an orbit about the Sun, and that on coming near the Earth they were promptly captured by it and caused to move about it as a satellite.’’  A few years later \citet{Denning1916} concluded that ‘‘the large meteors’’ that passed over Northern America in 1913 must have been temporary Earth satellites because they traveled 2,600 miles in the atmosphere suggesting that the orbits were ‘‘concentric, or nearly concentric, with the Earth’s surface.’’  Given that this event pre-dates the launch of any artificial objects it must have been a natural object and a minimoon by our definition.

\citet{Clark2016} suggest that a meteor observed on 2014 January 13 in the Czech Republic with an all-sky digital camera system that is part of the European Fireball Network has an $\sim95$\% probability of having been on a geocentric orbit before impact.  Complementary spectroscopic data prove that it must have been a natural object.  Detailed modeling of the object's atmospheric deceleration and fragmentation suggest that its pre-entry mass must have been about $5\kg$ with a diameter of $\sim15\cm$.  It entered Earth's atmosphere at just over $11.0\kps$, consistent with having a $v_{\infty}=0$ with respect to Earth as expected for geocentric objects, and their backward dynamical integrations suggest that it was a minimoon for at least $48\Day$ and perhaps for more than $5\yr$.  \citet{Clark2016} concluded that the predicted rate of minimoon meteors was far higher than the observed rate based on this object but we have confirmed that their estimated rate did not account for the vastly different detection efficiency of minimoon meteors compared to heliocentric meteors.  Meteor luminous efficiency, the fraction of a meteor's kinetic energy that is converted into visible light, is proportional to the 4\textsuperscript{th} or 5\textsuperscript{th} power of the impact speed so the apparent brightness of a meteor with a heliocentric origin \citep[$v\sim20\kps$;][]{Brown2013} will be $16\times$ to $32\times$ brighter than a minimoon meteor of the same initial mass.

\section{Minimoon dynamics}
\label{s.MinimoonDynamics}

\citet{Heppenheimer1977} defined `capture' as `the process whereby a body undergoes transition from heliocentric orbit to a planetocentric orbit'. Therefore, the three-body problem (3BP) is the natural framework to study the capture mechanisms for which the invariant manifolds of the orbits around the collinear Lagrange points are known to play a significant role. The capture definition entails that the body should remain gravitationally bound to the planet but, in a purely gravitational three-body scenario, captures can only be temporary \citep{Huang1983,Tanikawa1983}.

The three-body problem has no general analytical solution and is often simplified to the case in which two massive bodies are in circular orbits revolving around their centre of mass while the third body is massless and moving in their gravitational potential.  In this circular restricted 3BP (CR3BP) the dynamical system has an integral of motion that yields an invariant parameter known as the Jacobi constant, $C$.  It is related to the total energy of the particle in the synodic frame (the co-rotating frame with origin at the barycenter and the line between the two primary objects fixed) and its constancy imposes a dynamical constraint between the position and velocity of a particle. 

\begin{figure}[htbp]
\begin{center}
\begin{tabular} {ccc}
    \includegraphics[width=0.32\columnwidth]{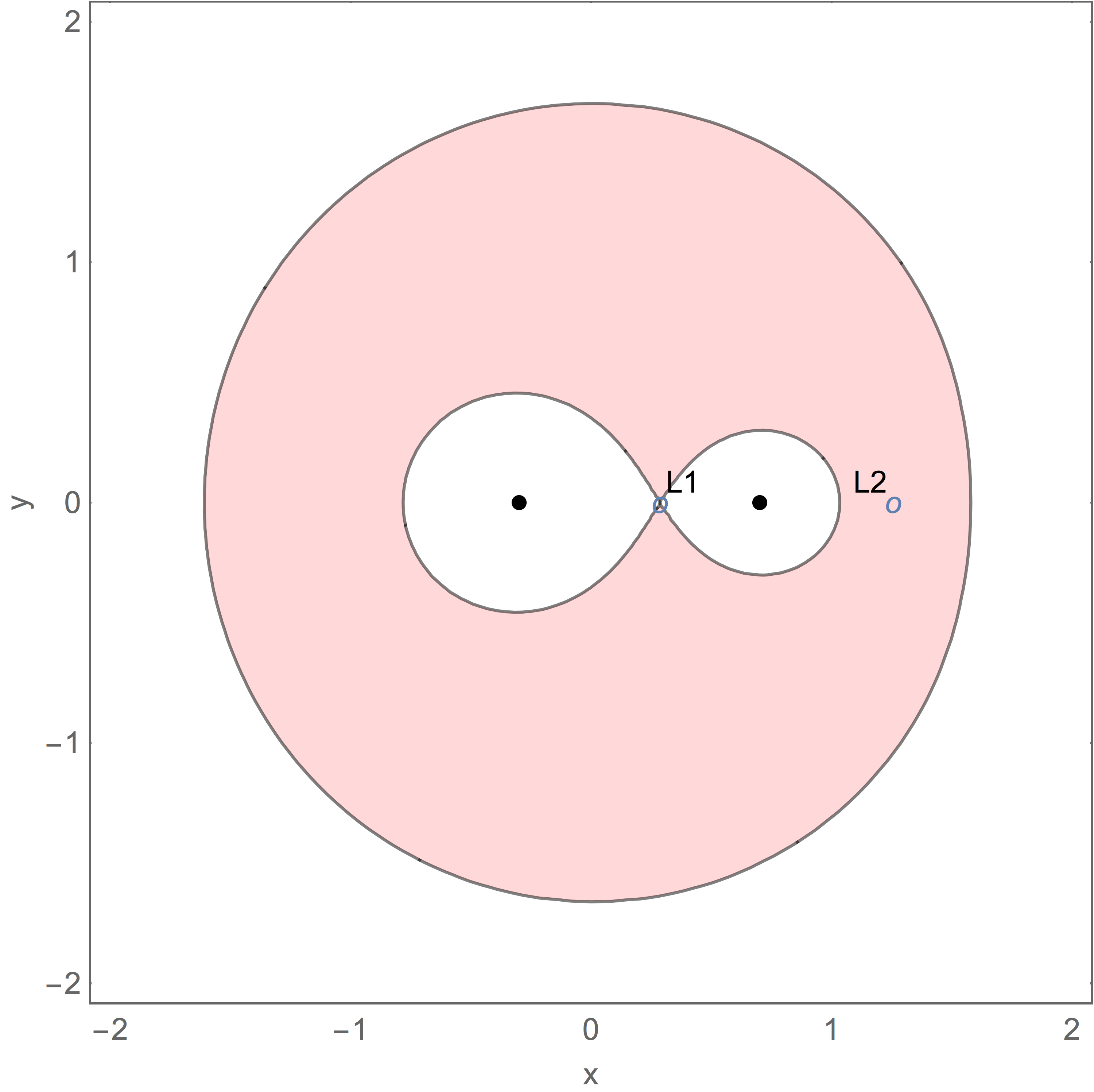} &
    \includegraphics[width=0.32\columnwidth]{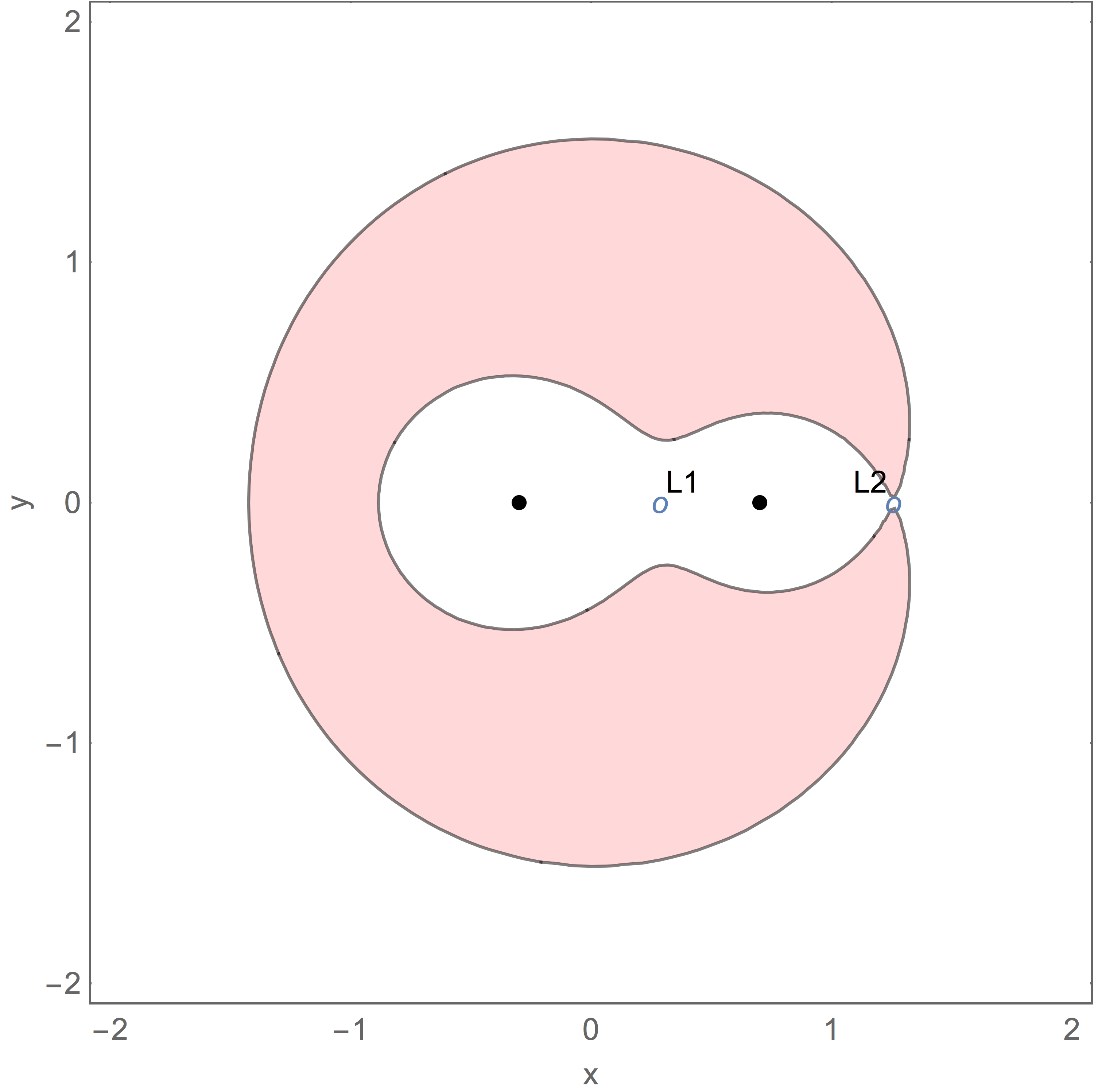} &
    \includegraphics[width=0.32\columnwidth]{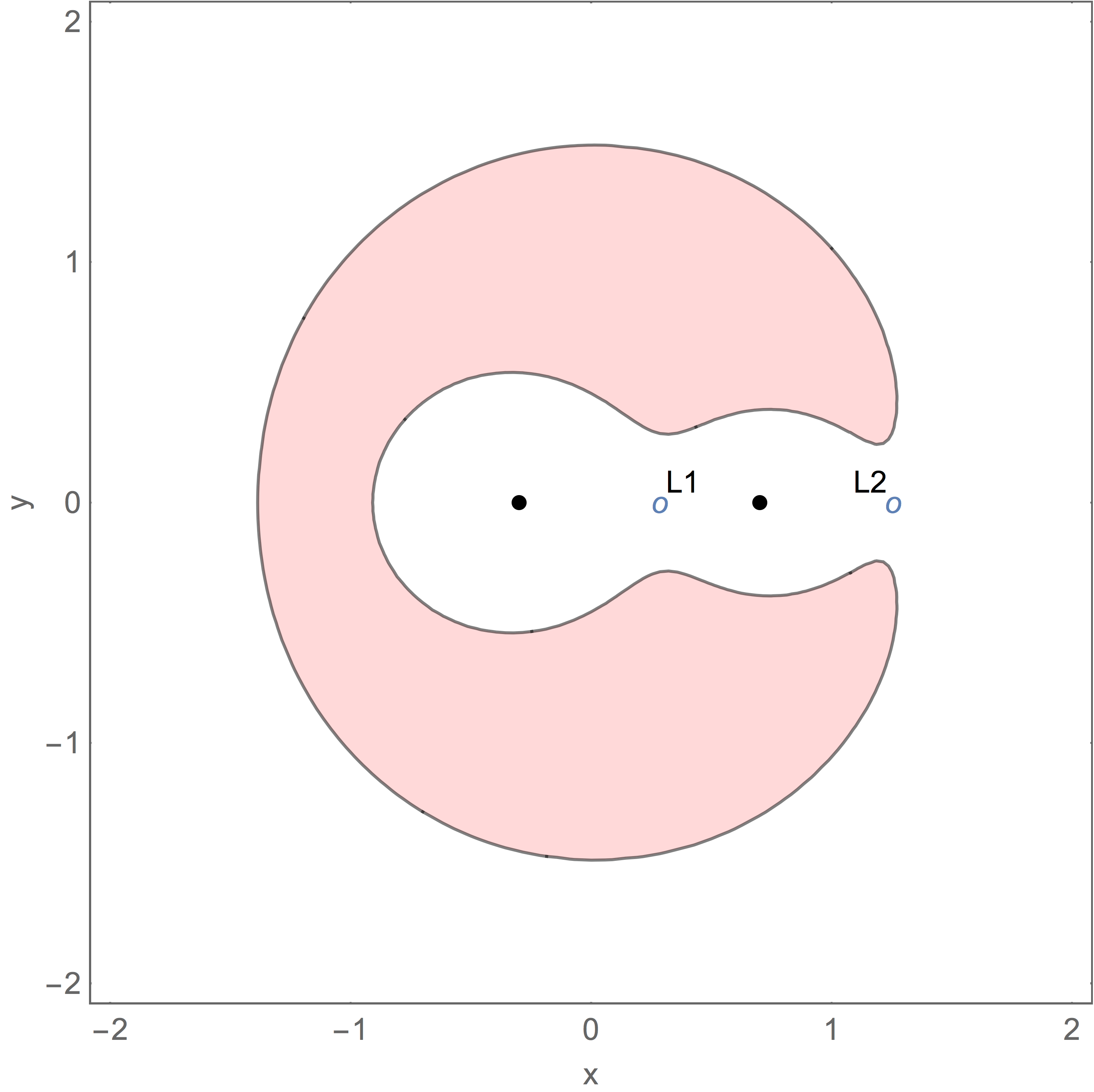} \\
    $C = C_1$ & $C = C_2$ & $C > C_2$
\end{tabular}
    \caption{Schematic view of the zero-velocity curves in the synodic frame for three different values of the Jacobi constant. The red shading illustrates regions where it is impossible for an object with the given value of the Jacobi constant to be located. The positions of the primary bodies are indicated by the filled black circles on the $y=0$ line while \Lone and \Ltwo are labelled and illustrated as unfilled circles.}
    \label{fig.ZeroVelocityCurves}
\end{center}
\end{figure}

For a given value of the Jacobi constant space is divided into \emph{forbidden} and \emph{allowable} regions (Hill regions) that are separated by  `zero-velocity' surfaces \citep{Szebehely1967}. These surfaces are defined in the synodic frame where they are invariant and symmetrical with respect to the $x-y$ plane in the CR3BP. The surfaces' intersection with the $x-y$ plane yields the zero-velocity curves (\fig{fig.ZeroVelocityCurves}).  $C_1$ and $C_2$ are the values of the Jacobi constant on the zero-velocity surface at the \Lone\ and \Ltwo\ libration points, respectively. For the Sun-Earth-asteroid system (but without loss of generality), when $C < C_1$ there are three disjointed Hill regions where the asteroid can reside: 1) in close proximity to Earth; 2) in the vicinity of the Sun; and 3) in the exterior domain that extends to infinity. None of these regions are connected, so an asteroid that resides in the Hill region surrounding Earth is gravitationally trapped and cannot escape into heliocentric orbit and vice versa. When $C = C_1$ the Hill regions around the Sun and Earth connect at \Lone, and for $C_1 < C < C_2$ a pathway exists around \Lone\ that allows an asteroid to transition from heliocentric to geocentric orbit. Equivalently, when $C > C_2$ another gateway opens at \Ltwo, connecting the exterior Hill region and enabling distant asteroids to transition to geocentric orbit. Hence, in the CR3BP framework it is impossible to effect a {\it permanent} capture because when the Jacobi constant is such that transfers from heliocentric to geocentric orbits are allowed there is no way to prevent the asteroid from returning into heliocentric orbit. The capture and escape trajectories are both governed by manifold dynamics, so once asteroids reach the vicinity of \Lone\ or \Ltwo\  the invariant manifolds of libration orbits are able to attract and pull them into the region around the planet following a stable manifold where they remain temporarily captured until they escape following an unstable manifold \citep{Carusi1981, Koon2001}. Note, however, that the duration of the temporary capture can be arbitrarily long.

The eccentricity of the Earth's orbits can be accounted for within the framework of the elliptic restricted three-body problem (ER3BP). An immediate consequence is that the Jacobi constant ceases to be an invariant quantity of the system (\ie, it is no longer constant) and Hill regions, as well as zero-velocity surfaces, are not invariant either; instead, they become periodic, time-dependent functions. As the Earth revolves around the Sun, the instantaneous Jacobi constant modulates and the Lagrange points shift inwards and outwards. Accordingly, at every value of Earth's true anomaly a different set of \emph{pulsating} zero-velocity surfaces exist with shapes and dimensions that vary in time. Hence, it might happen that the capture paths through \Lone\ and \Ltwo\ always remain closed or open, or open and close periodically every orbital revolution, depending on the geometrical layout and instantaneous value of the Jacobi constant. As a consequence, the eccentricity of planetary orbits is insufficient to provide a feasible capture mechanism on its own. Even if Earth's orbital eccentricity might enable the transition into geocentric orbit of asteroids that could not otherwise have transitioned within the CR3BP \citep{Mako2004}, there is no instrument to prevent them from returning into the heliocentric domain; the very same pathways will periodically reopen, thus enabling the asteroid's eventual escape. Therefore, in the gravitational three-body problem no dynamical mechanism exists that enables permanent capture. Doing so requires dissipative mechanisms that produce an irreversible change in the value of the Jacobi constant so that an asteroid may enter geocentric orbit through an open gateway which later closes before the asteroid can escape. Such dissipative mechanisms can only appear through the action of non-gravitational forces \citep[\eg][]{Pollack1979, Astakhov2003}, or the introduction of other perturbing bodies \citep[\eg, other Solar System bodies,][]{Nesvorny2007-IrregularSatellites}.

The Earth's case is more complex due to the subtle dynamical implications of the Moon so that a reliable study of the temporary capture of Earth's minimoons needs to be addressed within the framework of the Sun-Earth-Moon-Asteroid four-body problem. 


Despite the evidence of the `Chant Procession', the minimoon \designation{2006}{RH$_{120}$}, and the well known properties of temporary captures of comets and asteroids by the Jovian planets \citep[\eg][]{Carusi1981,Ohtsuka2008-147P,Vieira-Neto2001}, the first estimate of the number and properties of the EMS's steady-state minimoon population was performed by \citet{Granvik2012}.  They generated a synthetic population of NEOs that are the minimoon `source' population --- the set of objects that may be captured in the EMS --- according to what was at that time the best estimate of the NEO orbit distribution \citep{Bottke2002}, and then used an N-body integrator to simulate their dynamical evolution and determine the fraction that would be captured in the EMS.  They included the gravitational effects of the Sun, Moon, Earth, and the seven other planets and found that about 0.00001\% of all NEOs are captured as minimoons (TCOs) per year (\ie\ $10^{-7}$ of the NEO population per year).  This may seem like an insignificant fraction but there are estimated to be on the order of $10^9$ NEOs larger than $1\meter$ diameter \citep[\eg][]{Brown2013,Schunova-Lilly2017}, implying that a population of small minimoons is possible.  
In their careful accounting of the capture probabilities \citet{Granvik2012} calculated that there are likely one or two minimoons $\gtrsim 1\meter$ diameter in the EMS and that there should also be a $\sim 10\times$ larger population of temporarily captured flybys (TCF).  The average minimoon spends about 9 months in our system during which it makes almost 3 revolutions around Earth.

\begin{figure}[htbp]
\begin{center}

    \includegraphics[width=0.9\columnwidth]{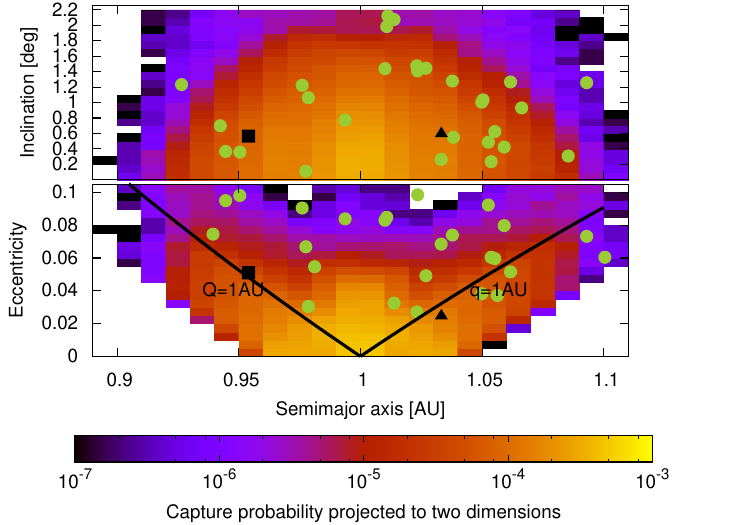}
    \caption{
      Combined TCO and TCF capture probability in heliocentric orbital element \aei\ space \citep[adapted from][]{Fedorets2017}. Green circles represent orbital elements of known NEOs as of 4 November 2014. Solid black lines correspond to $q=1\au$ and $Q=1\au$, perihelion and aphelion at Earth's orbit respectively. The black square represents the orbital elements of \RH\ at capture and the black triangle represents its current  orbital elements.
    \label{fig.MinimoonCaptureProbability}
    }

\end{center}
\end{figure}

\citet{Fedorets2017} improved upon the earlier work of \citet{Granvik2012} in a number of ways, notably by using an improved NEO model \citep{Granvik2016a-NEOModel-Nature} and a more careful accounting of the NEO orbital element distribution as $e \rightarrow 0$ and $i \rightarrow 0\arcdeg$.  The improved NEO model has a higher resolution in the orbital element distribution that was enabled by the use of much higher statistics and smaller time steps in the underlying dynamical integrations, and a significantly more careful analysis of the orbital element distribution of the main belt NEO `sources' (the main belt is the source of the NEOs just as the NEOs are, in turn, the minimoon source population).  Even with the higher resolution in the NEO orbital element distribution they found that it is still important to implement a more sophisticated treatment of the distribution of orbital elements {\it within} the bins at the smallest inclinations and eccentricities; \ie\ the \citet{Granvik2016a-NEOModel-Nature} NEO model specifies the number of objects in the bins that contain $e=0$ and $i=0\arcdeg$ but phase-space arguments suggest that the number distributions near zero should go as $n(e) \propto e^3$ and $n(i) \propto i^3$ \citep{Harris2016DPS-LifeNearZero}.  Since Earth-like minimoon pre-capture orbits are highly favored (\fig{fig.MinimoonCaptureProbability}) their improved treatment of the distribution caused a reduction of about $2\times$ in the predicted steady-state TCO population.  The reduction in the predicted TCO population was somewhat offset by a similarly more careful treatment of the TCF population.  Some of these objects may be bound to Earth for $>200\days$, they are more abundant than TCOs because of the reduced criteria for number of revolutions around Earth, and they have a slightly higher rate of impacting Earth during their capture.  Summarizing all their improvements, they found that the temporary natural satellite population (TCO+TCF) is smaller by $\sim10$\% compared to \citet{Granvik2012}'s estimate.

\citet{Urrutxua2017} subsequently refined the TCO and TCF definitions originally proposed by \citet{Granvik2012}.  They suggested that since temporary captures around Earth are best studied in a Sun-Earth synodic frame the number of revolutions should be counted by recording the angle swept by the ecliptic projection of the geocentric trajectory in the synodic frame. Accordingly, temporarily captured objects can be classified as TCOs when they complete at least one full revolution around  Earth or as TCFs if they fail to complete a full revolution under this definition.

\begin{figure}[htbp]
\begin{center}

    \includegraphics[width=0.45\columnwidth]{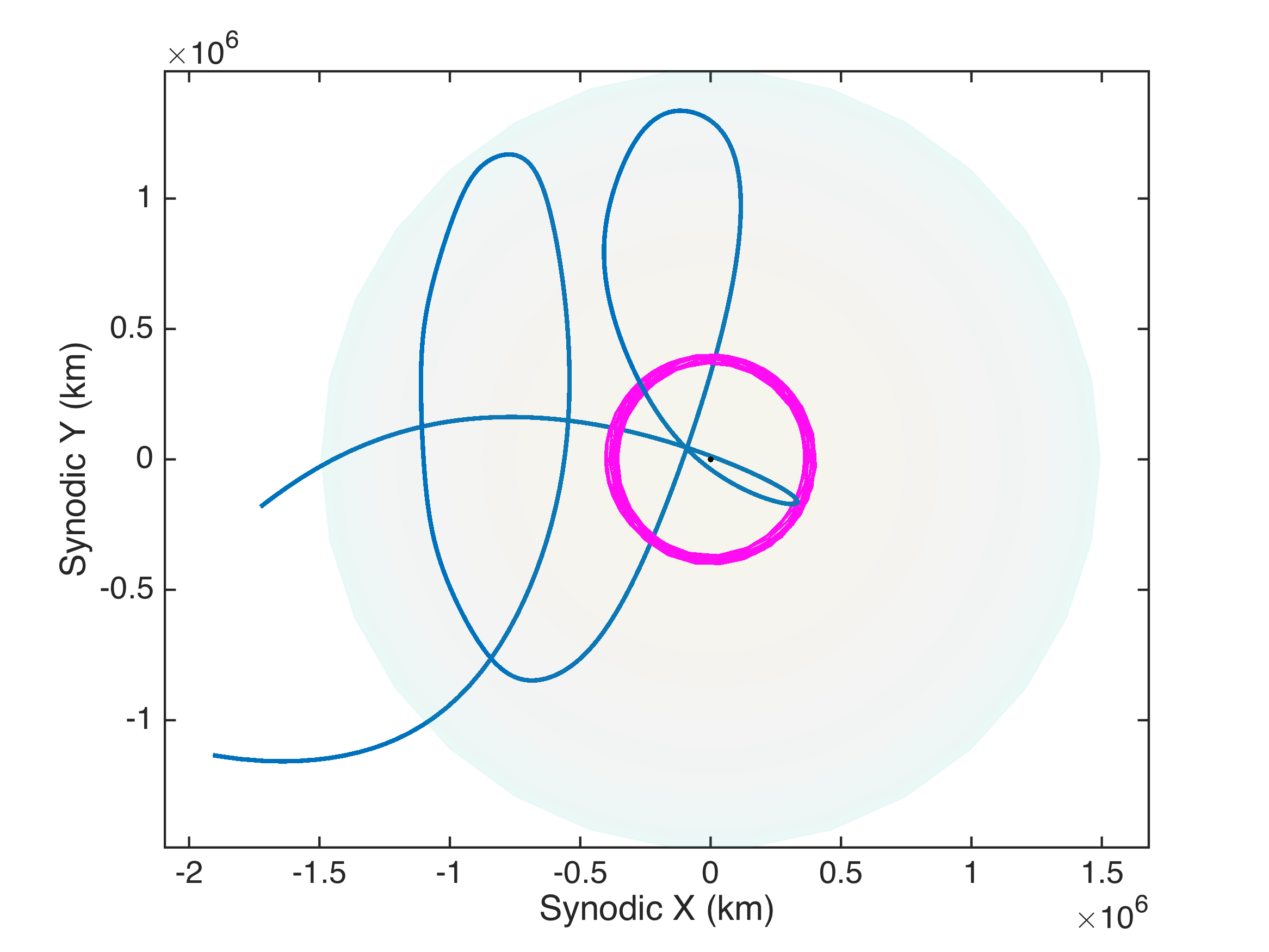}
    \includegraphics[width=0.45\columnwidth]{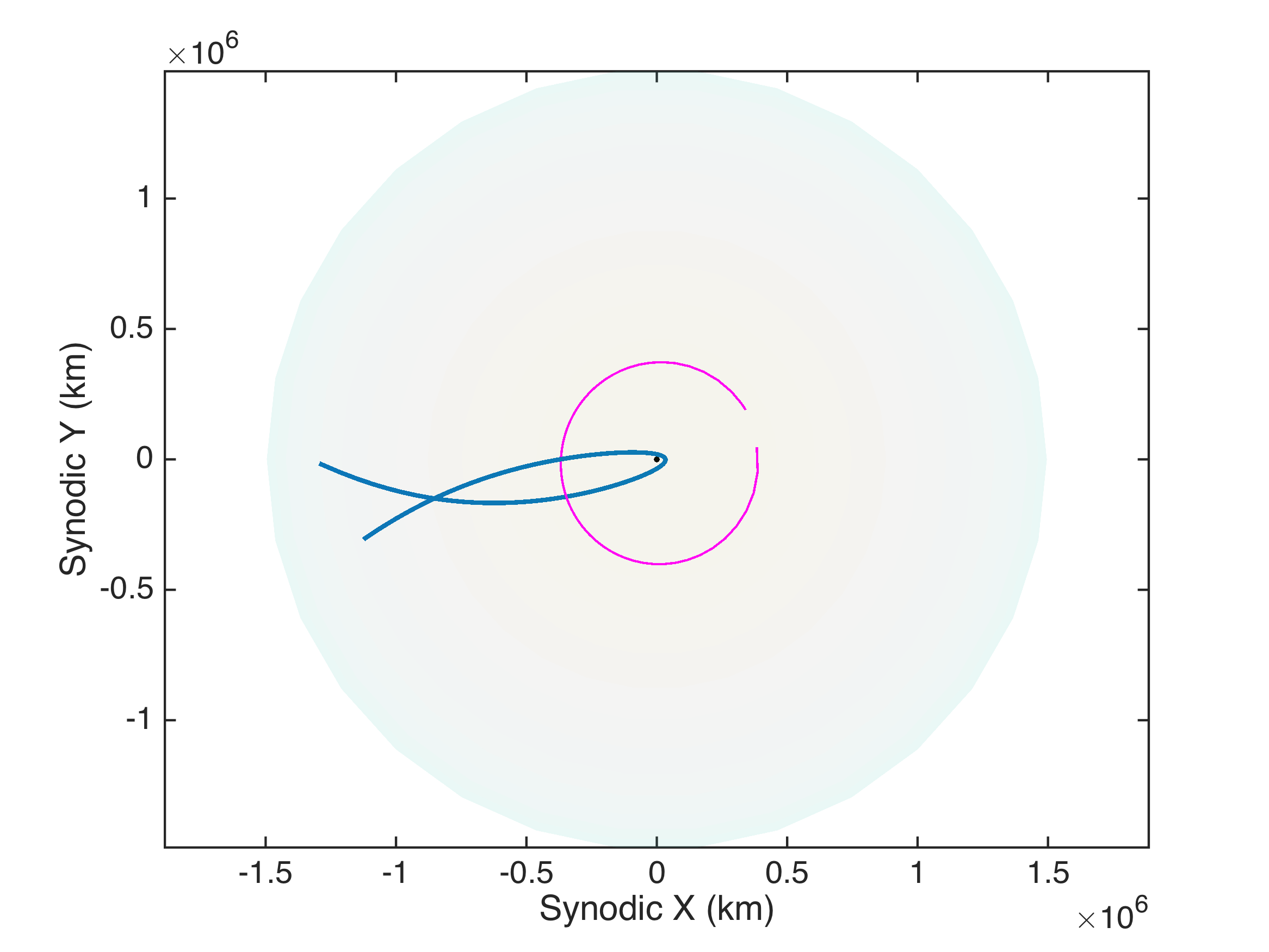}
    
    \caption{
    \citep[adapted from][]{Urrutxua2017}
    \textbf{Left:} Geocentric synodic trajectory of a TCF that becomes a TCO under the new definition of \citet{Urrutxua2017}. \textbf{Right:} A TCF that is misclassified as a TCO under the classical definition. The shaded area is the Hill sphere and the magenta curves depict the Moon's trajectory.
    }
    
    \label{fig.Misclassified-Minimoons}

\end{center}
\end{figure}

\begin{figure}[htbp]
\begin{center}

    \includegraphics[width=0.45\columnwidth]{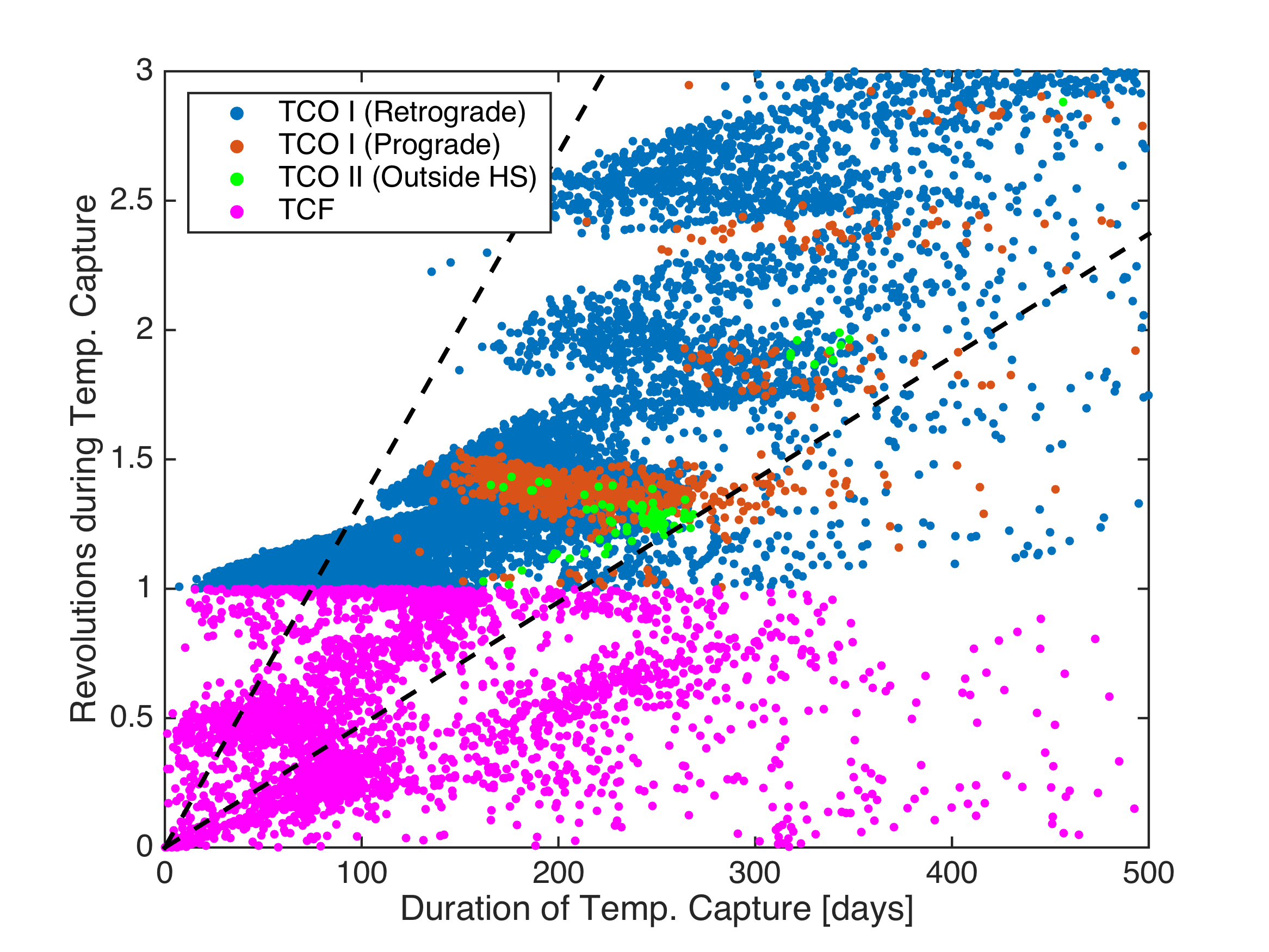}
    \includegraphics[width=0.45\columnwidth]{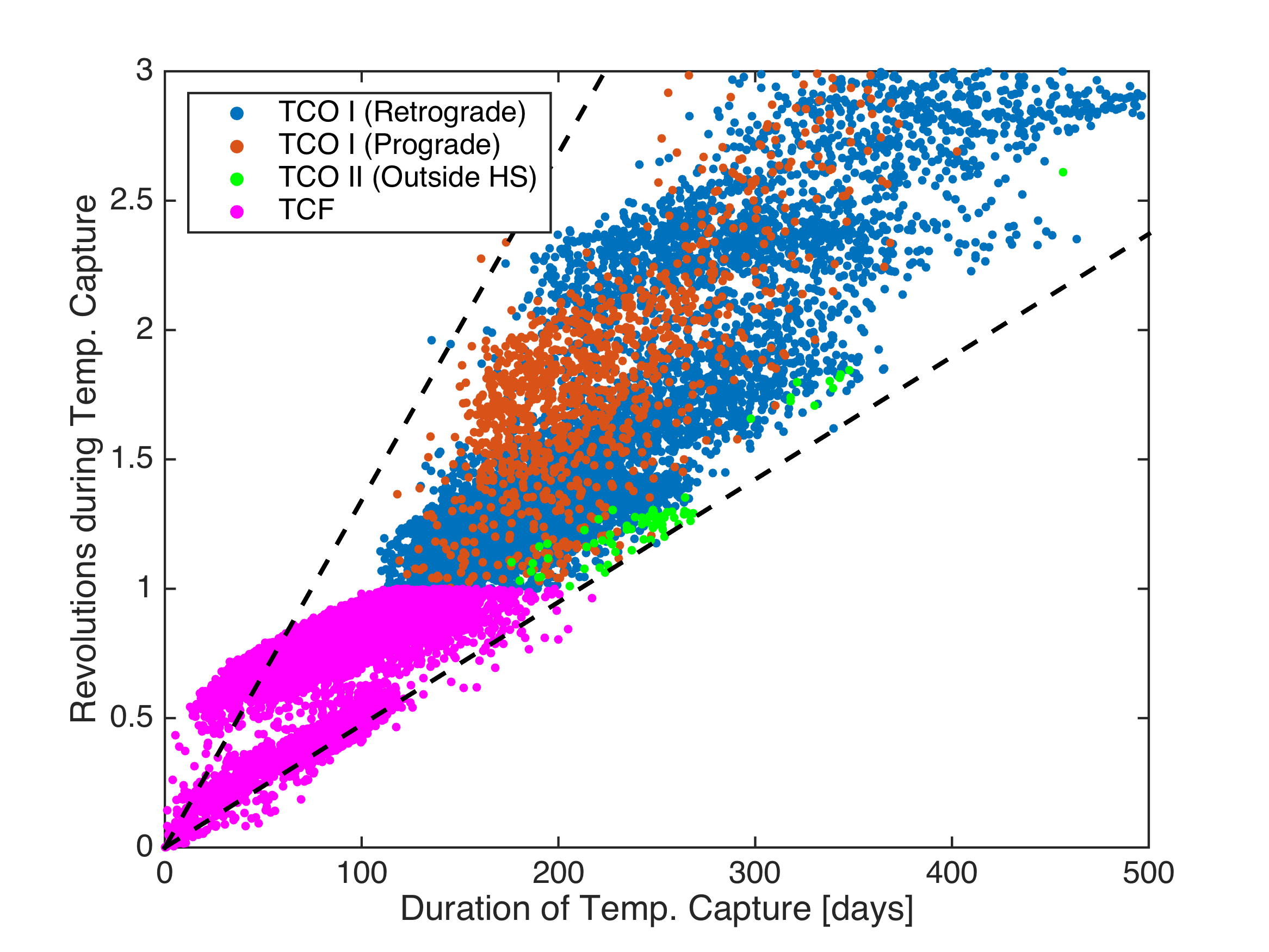}
    
    \caption{
    \citep[adapted from][]{Urrutxua2017}
    \textbf{Left:} TCO revolutions vs. capture duration for the definition and synthetic minimoon population of \citet{Granvik2012}. \textbf{Right:} The same population but using the definitions and method of revolution counting of \citet{Urrutxua2017}. Type I TCOs, both retrograde (blue) and prograde (orange), enter Earth's Hill sphere. Type II TCOs (green) remain outside the Hill sphere during their entire capture phase. TCFs (magenta) make less than one revolution around Earth (but the method for counting revolutions is different in the two panels). Dashed lines correspond to circular orbits at geocentric distances of 0.5 and 1~Hill radii.
    }
    
    \label{fig.Urrutxua.TCORevolutionsVsDuration}

\end{center}
\end{figure}

If TCOs followed circular orbits around Earth then there would be a linear  correlation between capture duration and revolutions with a different slope for each geocentric distance (\fig{fig.Urrutxua.TCORevolutionsVsDuration}).  The spread in the capture duration is thus linked to each TCO's average geocentric distance.  Although \citet{Granvik2012}'s minimoon sub-classification criteria is conceptually sound, unanticipated complications arise in practice. For instance, the synthetic minimoon in the left panel of \fig{fig.Misclassified-Minimoons} completes several `loops' during a temporary capture spanning 11~months though it only counts 0.93 revolutions about Earth and would be classified as a TCF according to \citet{Granvik2012}'s definition. Similarly, the synthetic minimoon in the right panel of \fig{fig.Misclassified-Minimoons} is bound within the EMS for barely a month while describing a short arc around Earth, yet the ecliptic projection of the trajectory happens to make more than one revolution so the object would be considered a TCO by \citet{Granvik2012}. These examples are contrary to common sense that would suggest that the TCF would be better classified as a TCO, while the TCO should be a TCF, \ie\ they appear to be misclassified. Examples of misclassified synthetic temporary captures are common, which indicated that the minimoon categorization algorithm required revision.

To address these issues \citet{Urrutxua2017} proposed the simple yet effective idea of counting the revolutions based on the intrinsic curvature of the synodic trajectory which is better suited to the three-dimensional non-elliptical nature of a minimoon's trajectory. It also decouples the definition from a geocentric reference and tracks the actual trajectory and the traversed arclength so it is more tightly linked to the dynamics and yields a stronger correlation between the capture duration and the number of completed revolutions (\fig{fig.Urrutxua.TCORevolutionsVsDuration}). The revised definition correctly reclassifies short-lived TCOs as TCFs, and long-lived TCOs with a previously small revolution count now have an appropriately higher number of revolutions. Thus, the `banding' in \fig{fig.Urrutxua.TCORevolutionsVsDuration} (left) is caused by TCOs whose synodic trajectories projected on the ecliptic describe loops that do not sum to the revolutions count under the classical definition (\eg\  left panel in \fig{fig.Misclassified-Minimoons}).

\citet{Urrutxua2017} also propose a classification scheme for TCO sub-types  (\fig{fig.Urrutxua.TCORevolutionsVsDuration}).  Type I TCOs cross the Hill sphere and are separated into retrograde and prograde orbits which reveals that, for an equal number of revolutions, prograde TCOs typically have  shorter capture durations than retrograde ones \ie\ the average geocentric distance during capture tends to be smaller for prograde TCOs.  Type II TCOs remain outside the Hill sphere and are long duration captures at any revolution count.

\begin{figure}[htbp]
\begin{center}
    \includegraphics[width=0.45\columnwidth]{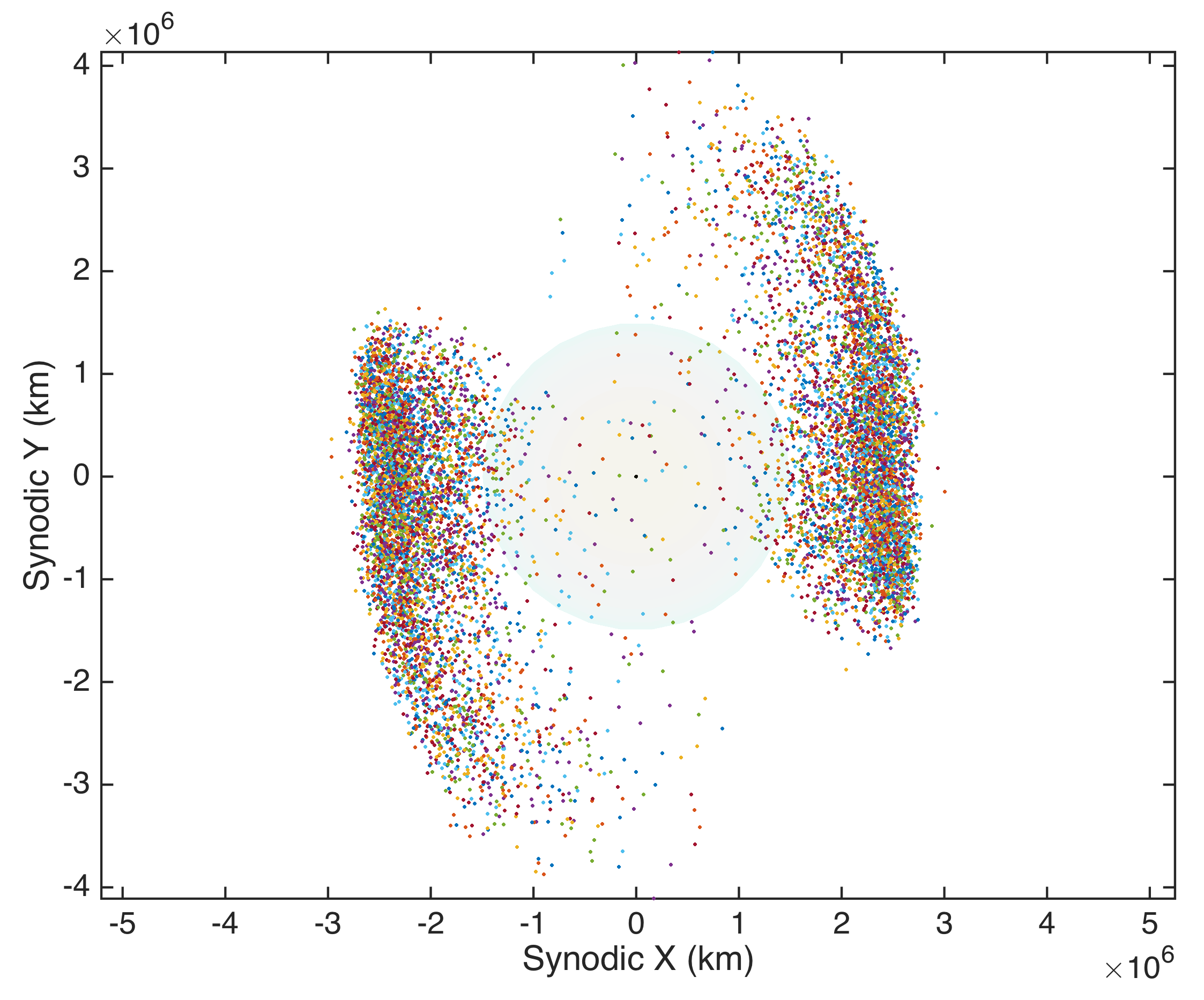}
    \caption{
      \citep[adapted from][]{Urrutxua2017}
      TCO capture location in the synodic frame at the moment that their energy becomes negative with respect to the Earth-Moon barycenter.  The Earth is located at the origin, the Sun is far off to the left, and the shaded grey circle represents Earth's Hill sphere.  There is no significance to the colors of the dots.}
    \label{fig.Urrutxua.TCOCaptureLocation}
\end{center}
\end{figure}

As described above, TCOs and TCFs are typically `captured' (\fig{fig.Urrutxua.TCOCaptureLocation}), \ie\ the moment their geocentric orbital energy becomes negative (\eqn{eqn.SpecificOrbitalEnergy}), when they are near the Earth-Sun \Lone\ or \Ltwo\ points \citep{Granvik2012}.  Their geocentric inclinations favor retrograde orbits in a 2:1 ratio, typical of irregular satellites and, perhaps surprisingly, the Moon has little to do with the capture process.  \citet{Granvik2012} established that the Moon is not important by running integrations with and without the Moon (but incorporating the Moon's mass into Earth) and found essentially identical capture rates from the NEO population.  The only significant difference was their finding that the Moon is a harsh mistress --- it causes TCO and TCF orbits to evolve to Earth-impacting trajectories while none of them impacted Earth without the Moon's influence.  There is no dynamical mechanism to shield Earth from minimoon impacts without the Moon but $\sim1$\% of minimoons strike Earth with the Moon in the simulation while $\lesssim0.02$\% of minimoons strike Earth without the Moon at the $90$\% confidence level.

Minimoon captures may begin over a wide range of geocentric distances (\fig{fig.Urrutxua.TCOCaptureLocation}) and, as noted above, TCOs may or may not cross the Hill sphere at all during their temporary capture. There is a strong symmetry in the incoming TCO distribution at the time of capture far from the Hill sphere but by the time they cross it the symmetry is lost and they are evenly distributed over the Hill sphere's surface. This suggests that the Hill sphere is not an appropriate reference surface for the study of temporary captures \citep{Urrutxua2017}.

\citet{Granvik2012}'s prediction that some minimoons can strike Earth provides a means of testing the minimoon theory because they calculated that about 0.1\% of all Earth impactors are TCOs.  TCO meteors have a distinctive signature in that their atmospheric impact speed is $\sim11.18\pm0.02\kps$ --- essentially Earth's escape velocity or, equivalently, the speed at which an object would strike Earth if it started at infinity with zero speed with respect to Earth.  Heliocentric meteors have $v_\infty>0$ and therefore must impact with speeds $>11.19\kps$.  They have an average impact speed of $\sim20\kps$ \citep[\eg][]{Hunt2004-MeteorVelocity,Taylor1995-MeteorVelocity} but can range in speed anywhere from $11.19\kps$ to $72\kps$.  The problem is that meteor luminous efficiency (and the radar echo as well) is a very steep function of the impact speed, so detecting a slow-moving meteor requires that the object be particularly large to be detected.  Thus, \citet{Clark2016}'s detection of a meteor with an origin on a geocentric orbit confirms \citet{Granvik2012}'s prediction that such objects exist but can not be used to test the minimoon population's size-frequency distribution without a detailed understanding of the detection biases.

Conversely, \citet{Hills1997-GrazingMeteoroids} calculated the probability that an Earth-atmosphere-grazing meteoroid could be captured into a geocentric orbit due to the loss of kinetic energy during its atmospheric passage.  They suggested that the cross section for orbital capture is about 1/1,000\textsuperscript{th} that of objects striking Earth which implies that the time scale for atmospheric capture of a $1\meter$ diameter object is a few decades --- much longer than the capture time scale calculated by \citet{Granvik2012} and \citet{Fedorets2017}.  Furthermore, objects that are captured by atmospheric drag must dive back into the atmosphere on every subsequent orbit, thereby rapidly dissipating kinetic energy until they fall to Earth as slow meteors.  Given their infrequent capture and short residence times we expect that this mechanism can not be a major minimoon source.

A sub-set of the minimoon population is the particularly long-lived orbits associated with the Earth-Moon \Lfour\ and \Lfive\ Trojan regions \citep[\eg][]{Hou2015-EM-Trojans,Marzari2013-EM-Trojans}.  These objects are deep within Earth's Hill sphere and can have lifetimes even up to a million years \citep{Hou2015-EM-Trojans} if they have small inclinations and eccentricities, and decameter-scale objects would even be stable under the influence of the Yarkovsky effect \citep{Marzari2013-EM-Trojans}.  The problem is that even though minimoons in the E-M Trojan population have very long dynamical lifetimes they are not long compared to the age of the solar system.  Thus, any E-M Trojan minimoon population must be transient but capturing NEOs into this sub-population is even less likely than the less restrictive captures described by \citet{Granvik2012} and \citet{Fedorets2017}.  Furthermore, there has never been a discovery of an Earth-Moon Trojan in the decades of operations of modern NEO surveys or in targeted surveys \citep{Valdes1983-EM-Trojans}.  We were unable to identify any limits on the size of population in the existing literature even though the requirements to do so are modest by contemporary asteroid survey standards \citep{Hou2015-EM-Trojans}.  One possible issue is that their typical apparent rate of motion would be about the same speed as the Moon's, $\sim12\degperday$, which is quite fast and would cause trailing of the detected asteroids on the image plane during typical exposures.  We expect that the LSST \citep{Ivezic2008-LSST,Schwamb2018-LSST-roadmap} will either detect the first E-M Trojans or set a tight upper limit on their size-frequency distribution.

A missing component from minimoon population modeling is an accurate incorporation of the Yarkovsky and YORP effects, thermal radiation forces and torques that cause small objects to undergo semimajor axis drift and spin vector modifications, respectively, as a function of their spin, orbit, and material properties \citep[\eg][]{Bottke2006-Yarko+YORP-Review}. These tiny thermal forces are partly responsible for allowing many of these bodies to escape the main asteroid belt in the first place.  At present, it is unclear how the inclusion of Yarkovsky thermal drift forces into our models would modify the minimoon capture rate near Earth but we suspect it would not be by very much because the change in semimajor axis produced by the Yarkovsky effect is probably on the order of $0.001-0.01\au\Myr^{-1}$, very small when one considers that their source NEO population is strongly affected by planetary close encounters.  It is probable that for every proto-minimoon moved onto a trajectory where capture was possible via the Yarkovsky effect, another would be moved off such a trajectory. Ultimately, though, new models are needed to fully evaluate their importance.   

The heliocentric orbits after capture remain `capturable' during subsequent Earth encounters \citep[\fig{fig.MinimoonCaptureProbability} and][]{Granvik2012}. This implies that artificial objects launched from Earth that escape the EMS to a heliocentric orbit can be captured during subsequent EMS encounters; \eg\ a recently discovered object and candidate minimoon, \designation{2018}{AV$_2$}, was initially predicted to have had an earlier capture in the late 1980s but follow-up astrometry later showed that a capture did not happen and that the object is likely artificial.  (It is nearly impossible to distinguish between minimoons and artificial objects based only on their orbital elements and dynamics but \S\ref{s.MinimoonCurrent+FutureStatus} describes how they can be differentiated using their response to radiative forces to measure their area-to-mass ratio.)

Finally, Earth is not the only world with minimoons. The most commonly known `minimoons' in the Solar System are associated with Jupiter whose Hill sphere is much larger than Earth's. Jupiter-family comets that evolve onto low-eccentricity, low-inclination heliocentric orbits similar to that of Jupiter can be captured in the Jupiter system via its \Lone\ or \Ltwo\ Lagrange points; \ie\ they form in the exactly the same way as described above for Earth's minimoons.  The most famous example was comet Shoemaker-Levy 9 that was likely captured around 1929 \citep{Chodas1996-SL9} and orbited within Jupiter's Hill sphere until it passed  within Jupiter's Roche limit.  This deep encounter disrupted the comet and created the famous ``string of pearls" that later returned to strike Jupiter in 1994. Other known comets have minimoon orbits with Jupiter \citep[\eg\ Comet 147P/Kushida-Muramatsu;][]{Ohtsuka2008-147P} but the steady state population has yet to be quantified with the latest dynamical models. Note that the orbits of Jupiter minimoons are different from Jupiter's irregular satellites, a population that exists on stable orbits with semimajor axes between 0.1 and 0.5 Jupiter Hill radii.  The irregular satellites were likely captured during a time of giant planet instability and migration that took place 4-4.5~Gyr ago \citep[\eg][]{Nesvorny2007-IrregularSatellites,Nesvorny2014-IrregularSatellites}.

\section{Minimoon source population}
\label{s.MinimoonSourcePopulation}

The minimoon source population, the set of objects from which minimoons are drawn, are Earth's co-orbital asteroids \citep{Morais2002-EarthCoOrbitals}, objects that are in a 1:1 mean-motion resonance with Earth like \designation{2010}{TK$_7$} \citep{Connors2011-EarthTrojan}, or at least those objects very close to the 1:1 mean-motion resonance, \citep[][]{Granvik2012,FuenteMarcos2013,Fedorets2017}. Thus, understanding the dynamics and population of co-orbitals is important to our understanding of the minimoon population as well.  The small population of known co-orbitals are all transient objects and therefore must not be primordial, having originated within the inner solar system, perhaps as impact ejecta from Venus, Earth, the Moon, or Mars, or, more likely, were delivered to the inner solar system from the main belt \citep[\eg][]{Granvik2017-escapeFromMainBelt}.

An interesting sub-class of asteroids that are tangentially related to minimoons are `quasi-satellites' \citep[\eg][]{Sidorenko2014-QuasiSatellites,FuenteMarcos2016-HO3,Chodas2016DPS-HO3}. Unlike geocentric minimoon orbits, quasi-satellites are heliocentric but their specific orbit elements while in the 1:1 mean-motion resonance cause them to appear to be in a distant retrograde orbit around Earth {\it from Earth's perspective}  (\fig{fig.Minimoon-vs-QuasiSatellite}).  This type of orbit can be dynamically stable because they never approach too close to any massive object and have been proposed for astrophysical and asteroid survey spacecraft missions because they provide inter-planetary-scale observations of Earth but at relatively constant geocentric distances \citep[\eg][]{Perozzi2017-DRO,Stramacchia2016-DRO,Cyr2000-SpaceWeatherDiamond}.  

Like minimoons, quasi-satellites are not just dynamical mathematical curiosities --- several examples are known to exist including asteroids (164207), (277810), \designation{2013}{LX$_{28}$}, \designation{2014}{OL$_{339}$}, and \designation{2016}{HO$_3$} \citep{FuenteMarcos2016-HO3,Chodas2016DPS-HO3}.  Both minimoons and quasi-satellites are drawn from the same NEO population and should have similar taxonomic distributions. However, the dynamical lifetimes of quasi-satellites can be orders of magnitude longer than for minimoons so it is to be expected that there should be more quasi-satellites and that the population should include larger bodies. The largest object that may be in the steady-state population at any time is directly related to the population lifetime; \eg\ the largest minimoon in the steady-state population at any time is likely $\sim1\meter$ diameter. Thus, given their long lifetimes , it is not surprising that quasi-satellites like \designation{2016}{HO$_3$} exist with an absolute magnitude $H\sim24.2$ corresponding to a diameter of $\sim50\meter$.

\begin{figure}[htbp]
\begin{center}

    \includegraphics[width=0.9\columnwidth]{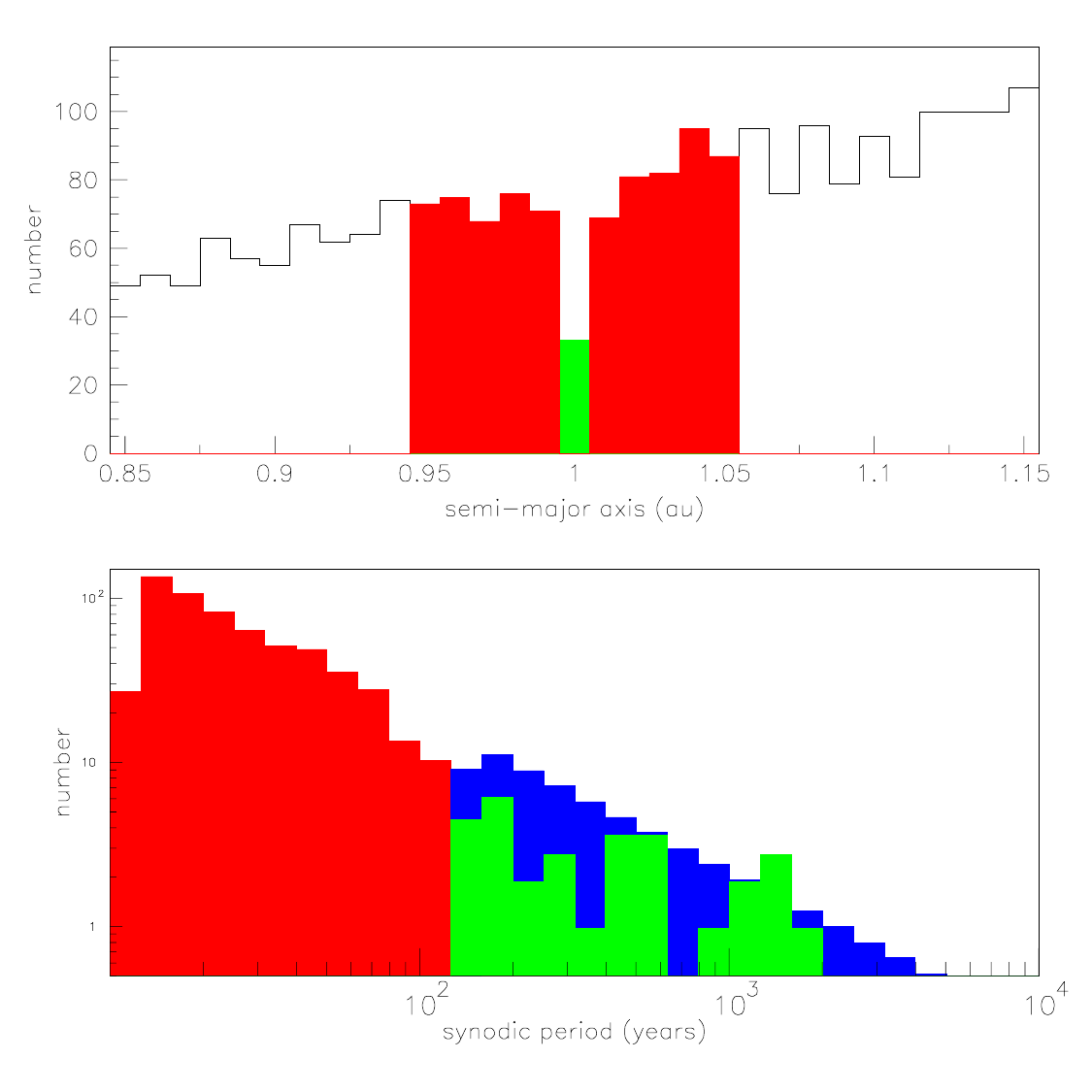}
    
    \caption{
      ({\bf top}) 
      The distribution of known NEO semi-major axes near $1\au$ as of 2018 Feb 25  (from astorb: \url{ftp://cdsarc.u-strasbg.fr/pub/cats/B/astorb/astorb.html}).  The green bins are for objects with semi-major axes very close to Earth's with $0.995\au < a < 1.005\au$, in or close to the 1:1 mean-motion resonance.  The red bins correspond to NEOs just outside that range with $0.945\au < a < 0.995\au$ and $1.005\au < a < 1.055\au$.
      ({\bf bottom}) The distribution of synodic periods color coded to the same objects in the top panel.  The blue histogram is the expected distribution of synodic periods if NEOs are distributed evenly in the range $[0.995\au,1.005\au]$ based on an extrapolation from the range $[0.845\au,1.155\au]$.
    }
    
    \label{fig.NEO-axis+synodic-near-1au}

\end{center}
\end{figure}

There is a clear lack of known NEOs with less than half the expected number of objects with semi-major axes within half a Hill radius of Earth's orbit (\fig{fig.NEO-axis+synodic-near-1au}).  We expect that this is an observational selection effect because NEOs in or near Earth's 1:1 mean-motion resonance have extremely long synodic periods (\fig{fig.NEO-axis+synodic-near-1au}).  The closer the NEO is to the 1:1 mean-motion resonance the longer its synodic period, making it much more difficult to discover.  Modern asteroid surveys have only been in operation for a couple decades so they have only an $\sim2$\% chance of detecting an NEO with a $1,000\yr$ synodic period.  Thus, the discovery of Earth's co-orbitals, and objects in the minimoon source population, simply requires a long period of time or more aggressive space-based observation platforms.

Finally, like minimoons, quasi-satellites are often touted as promising spacecraft mission targets because they are in not-too-deep space and always at relatively constant geocentric distances.  They are larger and easier to find than minimoons but require higher $\deltav$ and longer communication times and, since they are on orbits essentially identical to the minimoons' NEO source population, they will have the same taxonomic distribution as minimoons.

\section{Minimoon current status \& future prospects}
\label{s.MinimoonCurrent+FutureStatus}

The major problem with the minimoon hypothesis is the small number of known objects that have ever been minimoons (\S\ref{s.MinimoonDiscoveries}).  On the other hand, there have been numerous cases of objects that were TCOs or TCFs that later turned out to be artificial objects.  It would seem that the tremendous success of the current generation of NEO surveys at finding different classes of objects throughout the solar system ranging from a nearby and fast interstellar object \citep[\eg][]{Meech2017-Oumuamua-Nature} to distant and slow scattered disk objects \citep[\eg][]{Chen2016-RetrogradeTNO} should translate into more minimoon discoveries.  To assist in identifying geocentric objects the JPL Scout system\footnote{\url{https://cneos.jpl.nasa.gov/scout/intro.html}} \citep{Farnocchia2016-DPS-Scout} includes a geocentric orbit `score' to indicate whether an object may be bound in the EMS and it has been successful at properly recognizing artificial geocentric objects, particularly those with large semi-major axis.  So why haven't the surveys found more minimoons?

The explanation is simply that most minimoons are very difficult to detect.  \citet{Fedorets2017} calculated that the largest object in the steady-state population is likely only about $80\cm$ in diameter and the most probable distance is about 4 lunar distances or $0.01\au$ (a function of the orbit distribution and because the objects spend much more time at apogee than perigee).  At that distance a $1\meter$ diameter ($H \sim 32.75$) object at opposition has an apparent magnitude of $V \sim 22.7$ --- one magnitude fainter than the \PSone\ limiting magnitude for detecting main belt asteroids in its most efficient wide-band filter \citep{Denneau2013}.  Since minimoons will typically be moving much faster than main belt asteroids ($\sim3\degperday$ vs. $\sim0.25\degperday$) they will be more difficult to detect because their images will be trailed by more than the system's typical point-spread function.  When minimoons are closer they are brighter but also moving much faster, conditions under which the matched-filter algorithm\footnote{The matched filter algorithm is also known as the `shift-and-stack' algorithm or `synthetic tracking' or `digital tracking'.} applied to high-speed, low-noise cameras should excel \citep[\eg][]{Gural2005-MatcherFilter-Spacewatch,Shao2014-SyntheticTracking,Heinze2015-DigitalTracking}.  The problem is that these cameras are still only available in small formats (\ie\ small field-of-view).  Thus, the discovery of the next minimoon with the current survey systems will likely be of the serendipitous capture of a few meter diameter object like \designation{2006}{RH$_{120}$}, an event that occurs on the order of once a decade \citep{Fedorets2017}.

\begin{figure}[htbp]
\begin{center}

  \includegraphics[width=0.9\columnwidth]{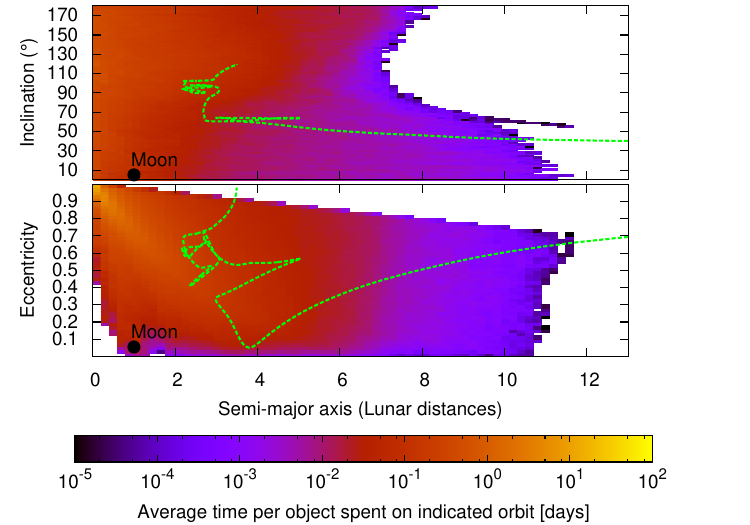}
  \caption{
    \citep[fig.~10a from][]{Fedorets2017}
    The residence time for synthetic minimoons as a function of their geocentric orbital elements. \ie\ the amount of time that minimoons spend on orbits with a given $\aei$ combination. The green dashed line is the trajectory of \RH\ through this representation of the orbital element phase space.
  }
    
  \label{fig.Minimoon-aei-longLivedOrbits}

\end{center}
\end{figure}

Even though minimoons and minimoon-like objects are difficult to detect the asteroid surveys do identify objects on a geocentric orbit.  Most are quickly associated with known artificial satellites but there are currently a few dozen unidentified geocentric objects\footnote{\url{https://www.projectpluto.com/pluto/mpecs/pseudo.htm}}.  Rapid follow-up on these objects is typically problematic because they are faint and have high apparent rates of motion.  As described above, these objects are usually dismissed as being artificial and this is probably true of almost all of them and especially so for the lower eccentricity, small revolution period objects.  However, the most likely minimoon geocentric orbits (\fig{fig.Minimoon-aei-longLivedOrbits}) overlap some of the longer period unidentified objects with high eccentricity.  Thus, while we agree that it is likely that most of the unidentified objects are artificial it should not be assumed that they are necessarily so. 

\citet{Bolin2014} performed an extensive analysis of existing capabilities for detecting minimoons and came to the same conclusion --- contemporary asteroid survey systems are only capable of serendipitous detections of the largest minimoons on decadal time scales.  They also explored options for fortuitous minimoon discoveries with existing space-based surveys such as NEOWISE \citep[\eg][]{Mainzer2011a} and with all-sky meteor surveys such as CAMS \citep{Jenniskens2011}, CAMO \citep{Weryk2013}, and ASGARD \citep{Brown2010} and, again, arrived at the conclusion that minimoon discoveries must be rare.  They suggested that targeted observations with a two-station (bi-modal) radar system would have a high probability of detecting a $>10\cm$ diameter minimoon in about $40\hours$ of operation but they note that their estimates are optimistic and that the effort may not justify the expense.  Their conclusion was that LSST could detect many minimoons and that a targeted multi-night survey with Hyper Suprime-Cam \citep[HSC;][]{Takada2010} on the Subaru telescope on Maunakea had a small chance of detecting a minimoon and would certainly be able to set a limit on the population statistics.

\citet{Jedicke2017-DPS-minimoonsJ} then obtained five nights of targeted minimoon surveying with HSC on Subaru under excellent conditions in an observing cadence specifically designed to identify geocentric objects over the course of a single night.  They acquired about 5 images of the same near-opposition fields spaced roughly evenly over about 4 to 6 hours in a field-of-regard of about 1,000$\degrees^2$ (\ie\ the total survey area).  They predict that they have about a 10\% chance of discovering a minimoon but the data analysis is still in progress.  Even without discovering a minimoon the data will allow the calculation of the first controlled upper limit on the minimoon population.

\begin{figure}[htbp]
\begin{center}

  \includegraphics[width=0.45\columnwidth]{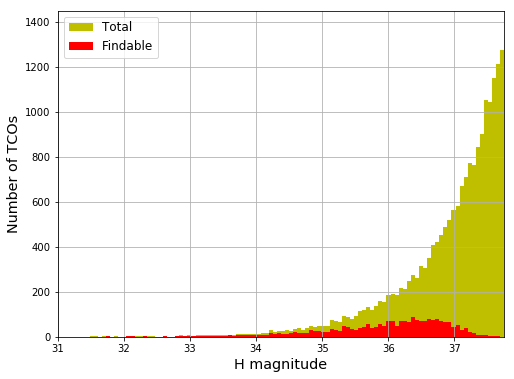}
  \includegraphics[width=0.45\columnwidth]{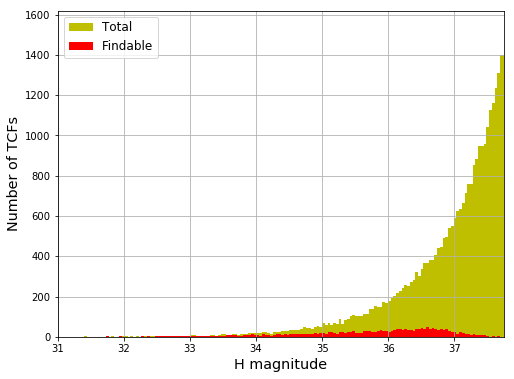}
    
  \caption{
    \citep[adapted from][]{Fedorets2015-LSST}
    Estimated number of TCOs ({\bf left}) and TCFs ({\bf right}) in the minimoon population (yellow) and which could be discovered in 10 years of LSST operations (red) as a function of absolute magnitude ($H$). A $1\meter$ diameter object has $H\sim33$ and a $10\cm$ diameter minimoon has $H\sim38$.
  }
    
  \label{fig.LSSTMinimoons}

\end{center}
\end{figure}

The LSST's advantages for minimoon discovery include its $8.4\meter$ diameter primary mirror that will achieve a limiting magnitude of $V\sim24.5$ in $30\second$ exposures over a $9.6\degrees^2$ field-of-view.  LSST is currently under construction on Cerro Pach\'{o}n, Chile and is scheduled to commence operations in 2022 \citep[\eg][]{Ivezic2008-LSST}.  \citet{Fedorets2015-LSST}'s simulated LSST survey was based on a current implementation of the expected survey pattern, weather, and performance characteristics to assess its performance for detecting minimoons.  The \citet{Fedorets2017} synthetic population of TCOs and TCFs was run through the LSST survey simulator and the output was then passed through their moving object processing system (MOPS) to emulate their baseline 10-year's of operations.  They found that LSST could discover many minimoons (\fig{fig.LSSTMinimoons}) and should efficiently and single-handedly discover\footnote{We use the word `discover' here to mean that LSST can detect the same minimoon multiple times in a single night {\it and} in at least three nights to determine its orbit.} essentially {\it all} the larger members of the population (if they can link detections of the same minimoon acquired on different nights). 

There remain at least a few difficulties with establishing the reality of new minimoons:  1) overcoming a prejudice against their existence, 2) obtaining evidence that they have a natural provenance and 3) establishing that they are not `merely` lunar fragments ejected from the Moon's surface during an impact event.  

The first issue will eventually be resolved when so many minimoons have been discovered that it is impossible to maintain a prejudice against them or when a serious flaw is discovered in the dynamical models that predict their existence.  

Resolving the second issue is a key input to the first but establishing the natural provenance of a tiny, fast moving, transient object is difficult (see the discussion on  \designation{2018}{AV$_2$} at the end of \S\ref{s.MinimoonDynamics}).  Apart from in-situ observations, the options for establishing a candidate as natural include obtaining spectra or colours, radar observations, or measuring its area-to-mass ratio (AMR) based on the magnitude of the effect of solar radiation pressure on its trajectory.  Obtaining sufficiently high signal-to-noise ratio (SNR) spectra of small, faint, fast objects is notoriously difficult and even low resolution colour photometry could require large telescopes and a disproportionate amount of observing time.  Radar observations can quickly establish an object's nature as the radar albedo easily differentiates between a natural rocky surface and the highly reflective surface of an artificial object, but there are few radar observatories in the world and it is not always possible to obtain radar observations of tiny, nearby objects that have very short round-trip times to the candidate; \ie\ minimoons are so close, and the reflected signal returns so fast, that they require bi-static observations in which one system transmits and the other receives.  Thus, perhaps the most straightforward manner of identifying natural objects is the AMR.  Artificial objects such as empty spacecraft booster stages or defunct satellites tend to have high AMRs while the few known small asteroids with measured AMRs are much smaller (\tab{tab.Area-to-Mass-Ratio}).  The typical minimoon candidate is so small that astrometric measurements over just a few month's time, comparable to the average minimoon's capture phase, have been sufficient to measure AMRs of similarly sized objects (\tab{tab.Area-to-Mass-Ratio}).  

\begin{table}[htp]
\caption{Area-to-Mass ratios (AMR) for select artificial satellites, the Moon, and small asteroids}
\begin{center}
\begin{tabular}{|c|c|c|c|}

\hline
object						    &	type		& 	AMR					            & reference  		\\
							    &			    &($\times10^{-4} \meter^2 \kg^{-1}$)& 			  	\\
\hline	
\hline	
Lageos 1 \& 2					&	artificial	&	$  7$				            & \citet{Beutler2006} \\
Starlette						& 	artificial	&	$ 10$				            & \citet{Beutler2006} \\
GPS (Block II)					&	artificial	&	$200$				            & \citet{Beutler2006} \\
\hline	
\designation{2006}{RH$_{120}$}	&	natural	    &	$ 11$	                        & ProjectPluto\tablefootnote{\url{https://www.projectpluto.com/pluto/mpecs/6r1.htm}}				\\
\designation{2009}{BD}			&	natural	    &	$  2.97 \pm 0.33$ 	            & \citet{Micheli2012-2009BD} \\
\designation{2011}{MD}		    &	natural	    &	$  7.9  \pm 7.4$	            & \citet{Mommert2014} \\
\designation{2012}{LA}			&	natural	    &   $  3.35 \pm 0.28$	            & \citet{Micheli2013-2012LA} \\
\designation{2012}{TC$_4$}		&	natural	    &   $  1.0  \pm 0.4$	            & JPL Small-Body Database\tablefootnote{\url{https://ssd.jpl.nasa.gov/sbdb.cgi?sstr=2012tc4}}	 \\
\designation{2015}{TC$_{25}$}	&	natural	    &   $  6-7 $	                    & \citet{Farnocchia2017-DPS-2015TC25} \\
Moon						    &	natural	    &	$  0.0000013$	                & \citet{Beutler2006} \\
\hline

\end{tabular}
\end{center}
\label{tab.Area-to-Mass-Ratio}
\end{table}

Having established that a minimoon is natural there still remains a `concern' that it could be fragment of lunar ejecta launched into geocentric or heliocentric orbit by the impact of a large asteroid on the Moon's surface.  We do not consider this issue to be of concern for many reasons.

First, the scientific and practical utility of a large piece of lunar ejecta is high; \eg\ for developing in-situ resource utilization technology and techniques.  A single $1\meter$ diameter lunar minimoon would have a mass of over $1,000\kg$ (assuming 50\% porosity and $5,000\kg\meter^{-3}$) while the six Apollo missions returned a total of about $382\kg$ or lunar material\footnote{\url{https://curator.jsc.nasa.gov/lunar/}} and the combined mass of all known lunar 
meteorites\footnote{\url{https://curator.jsc.nasa.gov/antmet/lmc/lunar_meteorites.cfm}} is about $65\kg$. While their is a tremendous scientific value associated with knowing the origin of the Apollo lunar samples it is also clear that lunar meteorites are important to our understanding of the Moon with 529 refereed journal papers listed on ADS\footnote{\url{http://adsabs.harvard.edu/}, The SAO/NASA Astrophysics Data System} including the words `lunar' and `meteorite' in the title.  We imagine that a verified lunar minimoon would have implications for the lunar cratering rate, impact ejecta models, dynamics in the EMS, measurement of Yarkovsky and YORP on small objects, \etc\  From an ISRU and human mission perspective it matters not whether a minimoon has a lunar or other origin as these objects provide small, low $\deltav$, cis-lunar candidates for testing system operations.

Second, \citet{Granvik2016a-NEOModel-Nature}'s dynamical simulations of orbital evolution of objects from the main belt into the NEO population show that there {\it are} dynamical pathways to the Earth-orbit-like heliocentric orbits necessary for capture in the EMS (\ie\ orbits with $a \sim 1\au$, $e \sim 0$, and $i \sim 0\arcdeg$).  Using that model, \citet{Fedorets2017} calculated that in the steady state there should be $3.5\pm1.4$ NEOs with $H<25$ on `capturable' orbits so there must be many more objects on those kinds of orbits at smaller sizes.  We stress that the NEO model already accounts for dynamical scattering of the NEOs by the EMS and should be considered the best possible model of the minimoon source population that is currently available.  (One possible issue is the impact of Yarkovsky on the evolution of the smallest NEOs as discussed earlier.)

Third, let's assume a large impact on the Moon took place, and that ejecta from this event delivered a large number of small objects from the Moon's surface to orbits within the Earth-Moon system.  Dynamical models suggest many will quickly impact Earth, the Moon, or will escape to heliocentric space.  For the latter, many may return at later times as impactors and potential minimoons. In this scenario, the impact capable of creating numerous meter-sized minimoons well after the event took place should also produce many lunar meteorites. Accordingly, we would predict that the petrology of many lunar meteorites should indicate they came from the same region while the cosmic ray exposure (CRE) ages of many lunar meteorites should have similar ages but neither prediction is supported by lunar meteorite studies.  \citet{Warren1994-LunarMeteorites} studied the delivery of lunar meteorites and argued that their formation craters are likely to have been both small and scattered across the Moon.  The CRE ages of lunar meteorites are consistent with this formation scenario as most of their ages are short ($<1\Myr$) with only a small fraction between $2-10\Myr$ \citep{Eugster2006-MESSII}.  There is little evidence for a group of lunar meteorites having similar ages. Note that the largest young impact crater on the Moon, the $22\km$ diameter Giordano Bruno crater, formed about $4\Myr$ ago yet there is no obvious indication that ejecta from this impact event is present in the lunar meteorite record. 
Accordingly, we are skeptical that lunar ejecta is a good source of present-day minimoons.  

\section{Minimoon science opportunities}
\label{s.MinimoonSciencesOpportunities}

Minimoons will provide interesting science opportunities as a consequence of their small sizes and their relatively long capture duration. Although similarly sized non-captured objects are much more numerous they are  typically observable for a much shorter period of time during their Earth fly-by. No meter-scale objects have ever been recovered during a subsequent apparition and hence their observability is limited to the discovery apparition. The minimoons' longer observation window allows for more detailed follow-up observations. In addition, the orbital uncertainty for minimoons becomes negligible  within a few days and therefore allows for detailed follow-up to be carried out earlier than for non-captured objects \citep[\fig{fig.MinimoonOrbitlUncertaintyEvolution} and][]{Granvik2013}.

The interior structure of meter-scale meteoroids is largely uncharted territory that could be tested with minimoons (it is arguable that the interior structure of asteroids of any size is largely unknown). There is essentially no data to constrain models that range from `sandcastles' held together by cohesive forces \citep{Sanchez2014} to solid, monolithic structures. Measured rotation rates are inconclusive because even small internal cohesive forces allow for faster rotation rates than would otherwise be possible for a non-rigid body. An asteroid's density provides some information to constrain its interior characteristics because we can assume that most of the material is `rocky' so a measured density less than rock implies that the interior contain voids \citep[\eg][]{Carry2012}. Asteroid volumes are typically based on photometry and/or radar data while mass estimation requires that it gravitationally perturbs a less massive test body such as a spacecraft or another much smaller asteroid \citep[\eg][and references therein]{Siltala2017}. Neither of these techniques is suitable for measuring a minimoon's mass but a minimoon's AMR (described above) can provide provide useful constraints on mass and density \citep[\eg][]{Micheli2012-2009BD,Micheli2013-2012LA,Mommert2014}. The AMR can provide a measure of an object's bulk density when combined with an estimate of its size and shape derived from lightcurve measurements. Minimoons, that spend months in Earth orbit, are particularly suited to AMR estimation since measuring the AMR requires that the object is small and tracked for a long period of time.


While remote minimoon measurements can be useful for answering some scientific questions we think it is clear that the most important science opportunities derive from in-situ minimoon measurements.  A small spacecraft mission could determine the shape and structure of a meteoroid, its regolith properties, and obtain high-resolution surface images in many wavelengths that can be compared to remote measurements of much larger asteroids.  Returning a minimoon to Earth will be difficult but minimoons could provide a tremendous amount of pristine asteroid material from many different asteroids.   Remember that meter-scale meteoroids deliver meteorites but only the strongest material survives passage through Earth's atmosphere, and impact and weathering on Earth's surface.  Minimoons provide an intact, pre-contact meteoroid in its entirety, with all the fragile components in their original context.

\section{Minimoon mission opportunities}
\label{s.MinimoonMissionOpportunities}

\begin{figure}[htbp]
\begin{center}

    \includegraphics[width=0.75\columnwidth]{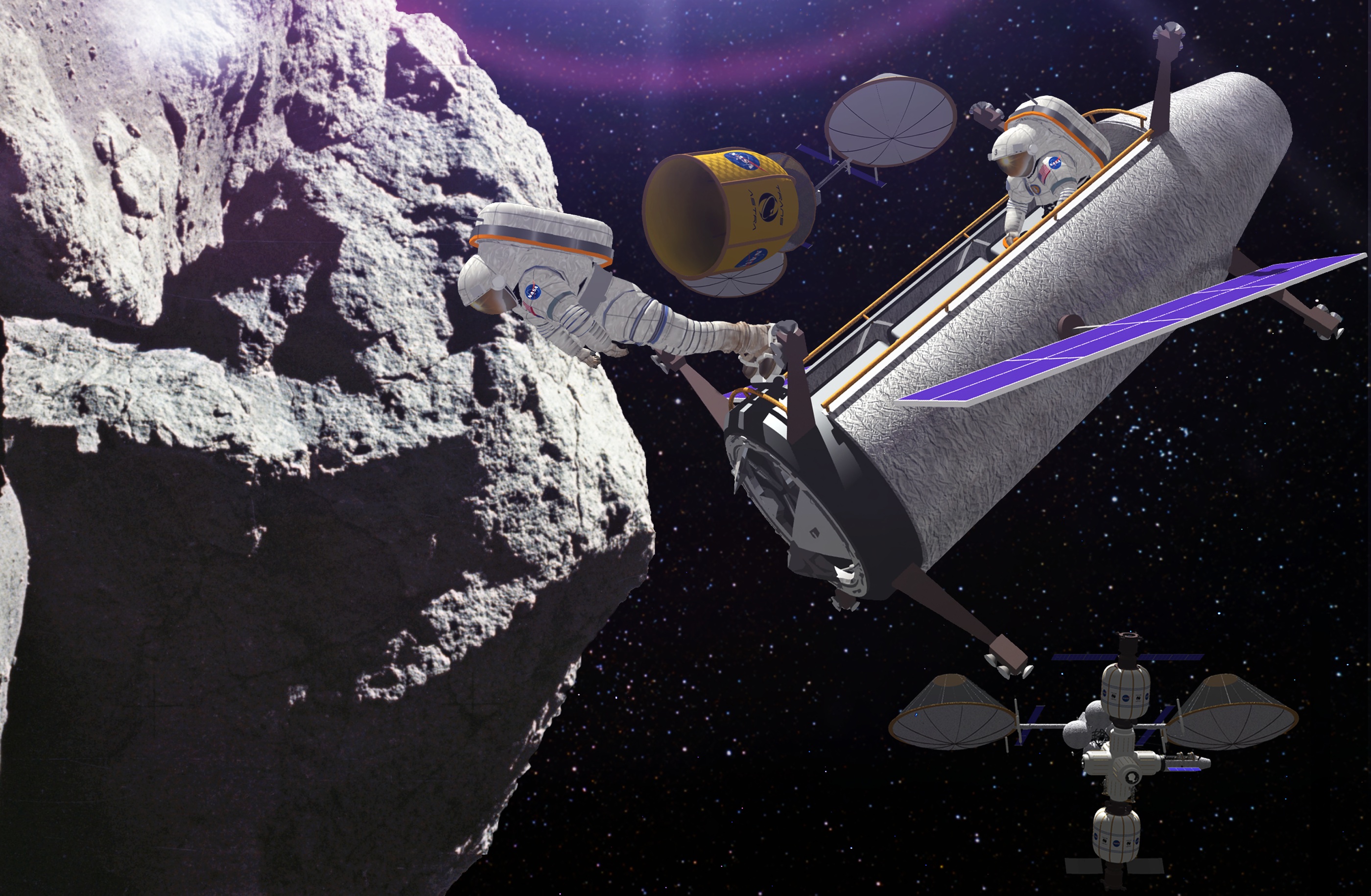}
    
    \caption{
      Artist’s illustration of asteroid ISRU showing astronauts at an asteroid as well as other mining and transportation vehicles operating in space (image credit:  TransAstra Corporation \& Anthony Longman).}
    
    \label{fig.TransAstraHumanExploration}

\end{center}
\end{figure}

After the discovery of \RH\ and the realization that there is likely a steady-state population of similar objects, Earth's minimoons have entered the game as candidates for future space missions.  They have been delivered for free to cis-lunar space by the solar system's gravitational dynamics and are now available in \emph{our own backyard} under  favourable energetic conditions which make them ideal targets.  Given their small size, Earth proximity, and their accessibility to long-term capture orbits, minimoons could enable affordable robotic and crewed missions using existing technology, as well as retrieval of substantially larger amounts of material compared to traditional sample return missions. Also, scaled versions of hazardous asteroid mitigation techniques could be tested at a fraction of the cost of current proposals. For all these reasons, minimoons stand out as compelling candidates for asteroid retrieval missions.

From a technological and commercial perspective they provide an ideal opportunity for: 1) the development and testing of planetary defence technologies (\eg\ deflecting an asteroid); 2) validating and improving close-proximity guidance, navigation, and control algorithms, 3) testing close-proximity procedures and protocols for safe operation of crewed missions around asteroids, and 4) establishing the feasibility of asteroid mining technologies for future commercial applications, all in an environment where the round-trip light-time delay is a few seconds. This short list illustrates that minimoons have far-reaching non-science implications for different stakeholders.

\begin{figure}[htbp]
\begin{center}

    \includegraphics[width=0.65\columnwidth]{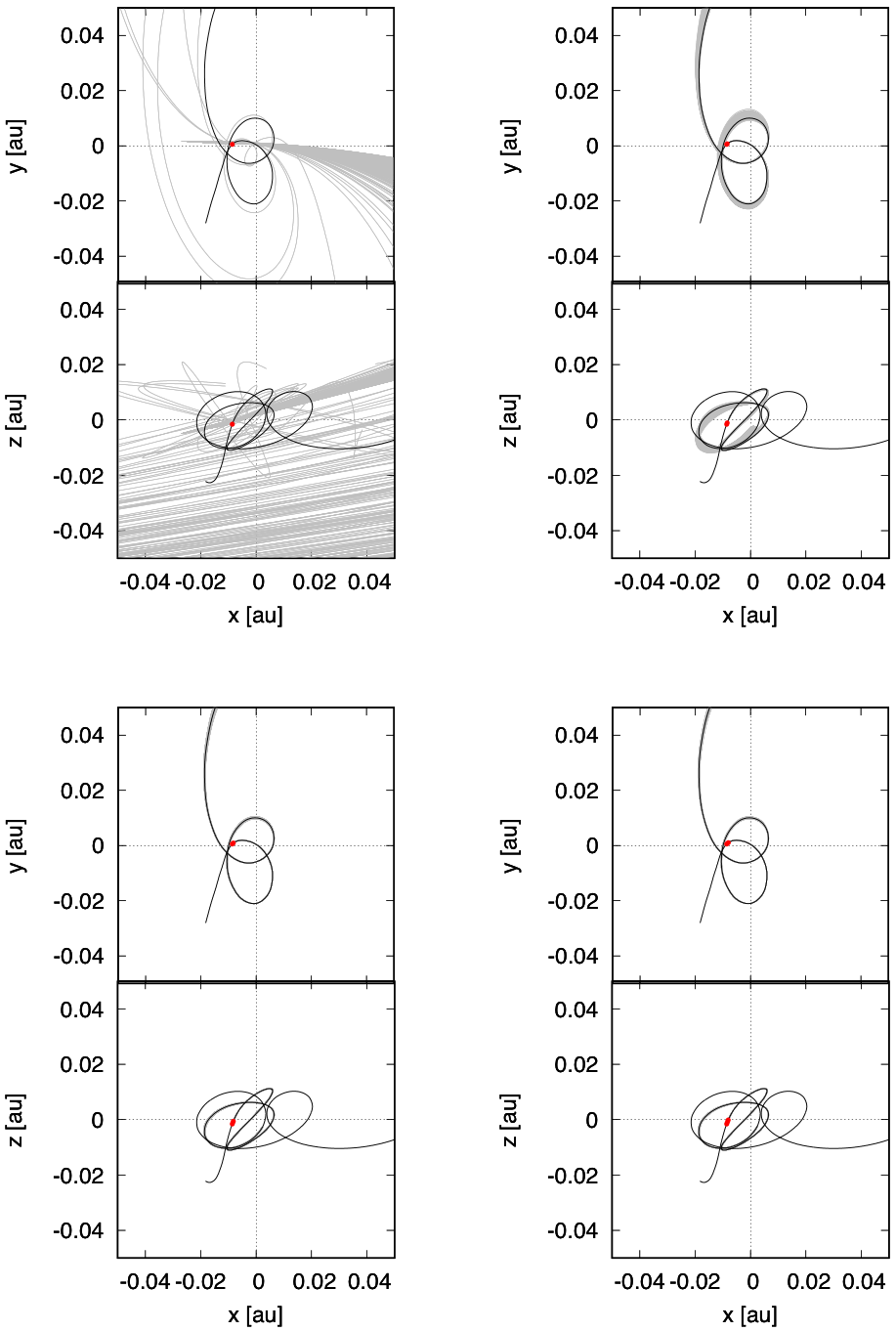}
    
    \caption{
      \citep[adapted from][]{Granvik2013}
      The evolution of the orbital uncertainty for a synthetic minimoon as a function of increasing observational timespan and number of observations; ({\bf top left}) 3 detections in one hour, ({\bf top right}) 6 detections in 25 hours, ({\bf bottom left}) 9 detections in 49 hours, and ({\bf bottom right}) 12 detections in 73 hours. The black line shows the true orbit in the $XY$ and $XZ$ planes in an ecliptic coordinate system that is co-rotating with the Sun so that the Earth is always in the center $(0,0,0)$ and the Sun is always at about $(1,0,0)$. The gray shaded area shows the extent of all acceptable orbits and the red dots mark the locations of the synthetic minimoon at the observation dates. All orbits were extended 500 days into the future starting from the date of the first observation.
    }
    
    \label{fig.MinimoonOrbitlUncertaintyEvolution}

\end{center}
\end{figure}

%
%

Many studies have suggested that a substantial amount of asteroidal resources can be accessed at an energy cost lower than that required to access resources from the Moon's surface \citep[\eg][]{Sanchez2011,Sanchez2013,Jedicke2018}.  Very simply, the lower the required $\deltav$ for a spacecraft to return from mining an asteroid, the lower the cost of the mission and, more importantly, the higher the profit.  Known NEOs are accessible with much lower $\deltav$ than main belt asteroids \citep[\eg][]{Elvis2011,Garcia2013,Taylor2018-MainBeltDeltaV} and the population of yet-to-be-discovered small NEOs on Earth-like orbits offers the possibility of many more commercially profitable asteroid missions \citep{Jedicke2018}.

These ideas has been around for a while in the realm of speculative science and science-fiction literature and have recently started to gain popularity in the public and private aerospace community. The renewed interest has led to the development of new trajectory designs, and asteroid retrieval and mining concepts \citep[\eg][]{Brophy2013,Strange2013-ARM,Graps2016,Sercel2017,Jedicke2017DDA}. Some of these technologies involve the artificial deflection of an asteroid's trajectory to shepherd it into cis-lunar space; \ie\ the creation of human-assisted natural minimoons \citep[][]{Garcia2013a, Chen2016a}.  In these scenarios, the selection of target asteroids is usually driven by minimizing a mission's $\deltav$ (cost). Naturally captured minimoons provide an excellent, easily-accessible testbed for developing those technologies \citep[][]{Granvik2013}. 

\citet{Baoyin2010} proposed capturing asteroids passing close to Earth by providing them with the necessary $\deltav$ so that zero-velocity surfaces would close within the framework of the CR3BP (\ie\ creating minimoons) and their best (known) target asteroid, \designation{2009}{BD}, only requires a $\deltav \sim 410\mps$.  \citet{Hasnain2012} then studied the total $\deltav$ required to transport an asteroid into Earth's sphere of influence including capture, concluding that a $\deltav = 700\mps$ for \designation{2007}{CB$_{27}$} was the best opportunity for a known asteroid. A lunar flyby can be used to provide some of the required $\deltav$ for capture in the EMS as shown by \citet{Gong2015} who obtained a long duration capture with a $\deltav = 49\mps$ for asteroid \designation{2008}{UA$_{202}$}.  It is important to note that all these studies were limited to {\it known} objects --- the number of objects increases dramatically at smaller sizes for which the known population is only a small fraction of the total population.  Thus, in the future, there will undoubtedly be many more objects available at even lower $\deltav$, especially if space-based missions are designed specifically to identify these targets.

In a search for novel minimoon capture-enhancement strategies, NASA developed an innovative mission concept to deliver asteroid \designation{2008}{HU$_4$} into a stable `distant retrograde orbit' (DRO) around the Moon (\ie, a minimoon on a geocentric orbit such that it becomes a quasi-satellite of the Moon in the EMS), with an estimated $\deltav \sim 170\mps$ \citep{Brophy2012}. DROs are stable solutions of the three-body problem that can be used whenever an object is required to remain in the neighborhood of a celestial body without being gravitationally bound \citep[\eg][]{Perozzi2017-DRO}. 

Another interesting strategy was proposed by \citet{Garcia2013}, who utilized the CR3BP invariant manifold dynamics to identify low energy asteroid retrieval transfers. In particular, they coined the term `Easily Retrievable Objects' to refer to the subclass of NEOs that can be gravitationally captured in bound periodic orbits around the Earth-Sun \Lone\ and \Ltwo\ points. Interestingly, the lowest $\deltav$ object was \RH, the first minimoon, that is now on a heliocentric orbit, at an astounding $\sim 50\mps$.

The utility of minimoons as spacecraft targets may be limited by the length of time they remain captured \citep[average capture durations of about 9 months;][]{Granvik2012,Fedorets2017} but there are at least two ways to overcome this limitation: 1) artificially extend the capture duration or 2) have rendezvous spacecraft emplaced and `hibernating' in a high geocentric orbit for serendipitous missions of opportunity once a desirable a minimoon is discovered.  Normal spacecraft-asteroid rendezvous mission time frames for proposal, development, launch, and operations are much longer than typical minimoon lifetimes and have not been considered to-date in the literature.

With the first vision in mind, \citet{Urrutxua2015} found that artificially extending a minimoon's capture duration could be accomplished in many cases at strikingly low $\deltav$s. They found that a $\deltav\sim44\mps$ (with slow deflection techniques) during \RH's minimoon phase in 2006-2007 could have extended its capture duration to over 5.5 years from its nominal 9 month's time in cis-lunar space.  In the unlikely scenario that the artificial deflection can begin {\it before} the temporary capture phase the authors concluded that by starting $\sim 316\days$ before perigee a total $\deltav\sim32\mps$ would have sufficed to extend the capture for an additional 5 years.  It might be argued that \RH\ was an unusual minimoon, so the authors extended their study to nine randomly selected virtual minimoons provided by \citet{Granvik2012} and found that some of their captures could be extended for decades at $\deltav$s of just $9\mps$. They also suggested that temporary captures could be artificially induced for asteroids that would otherwise not be captured at all and in-so-doing produce captures that last for decades with a small to moderate early deflection. Of course, the challenge  resides in identifying candidate asteroids with sufficient time to enable an asteroid retrieval mission to be planned and dispatched in a timely manner.

Several other studies suggest that capturing NEOs as minimoons is possible with small $\deltav$.  \citet{Tan2017-MomentumExchange} investigated opportunities using momentum exchange between an asteroid pair to capture one of the asteroids as the pair is directed close to one of the Sun–Earth \Lone\ or \Ltwo\ points.  They proposed the ambitious concept of first creating the asteroid pair by engineering a capture or impact during the fly-by of a small asteroid by a large one.  While their work shows that the process is possible, they note there remain ``significant practical challenges''.  The same three authors also examined less complicated ``direct capture'' mechanisms whereby the orbit of a heliocentric NEO is modified with a small $\deltav$ to induce capture in the EMS \citep{Tan2017-Direct+IndirectCapture}.  This scenario is essentially enhancing the natural minimoon capture process to capture specific NEOs onto long-lived geocentric orbits.  Similarly, \citet{Bao2015} studied the use of lunar and Earth gravity assists (LGA and EGA) in maneuvering NEOs into becoming minimoons.  They found that NEOs moving at $<1.8\kps$ with respect to Earth within Earth's Hill sphere could be captured using LGA and even higher speed objects could be captured using combinations of LGAs and EGAs.  The known NEO with the smallest capture $\deltav\sim76\kps$ is \designation{2000}{SG$_{344}$} but there are many, many more unknown NEOs that could be captured using these techniques.

\begin{figure}[htbp]
\begin{center}

    \includegraphics[width=0.30\columnwidth]{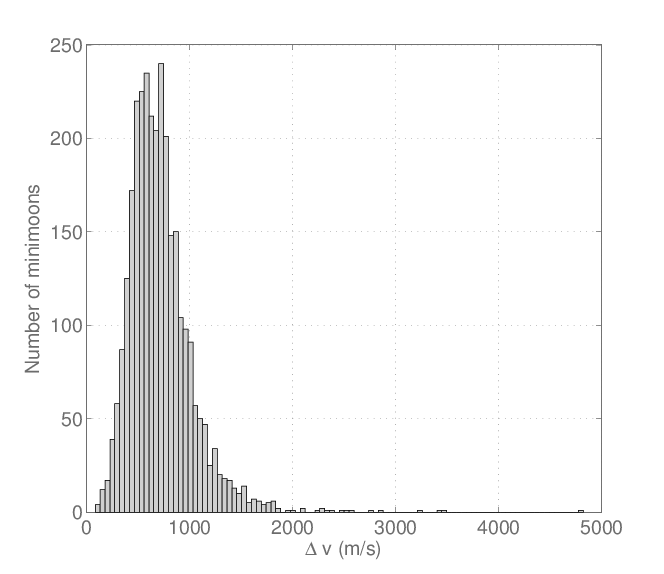}
    \includegraphics[width=0.5\columnwidth]{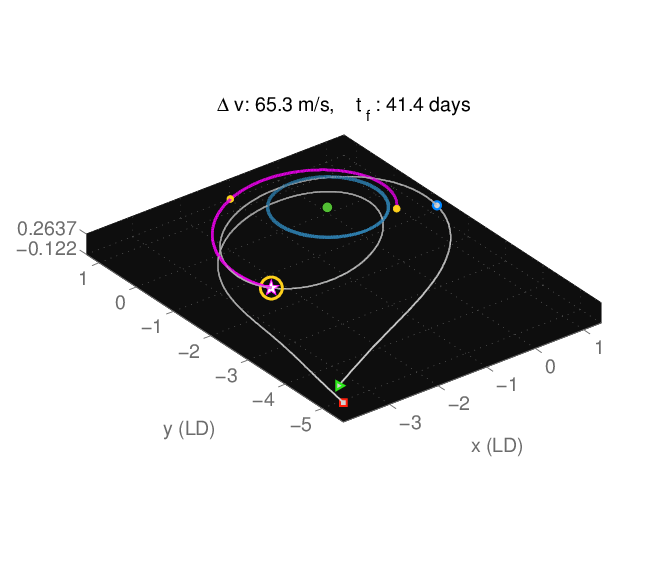}
    
    \caption{
      \citep[adapted from][]{Chyba2016}
      ({\bf left}) $\deltav$ distribution to 3,000 synthetic TCOs \citep{Granvik2012} from an Earth-Moon \Ltwo\ halo orbit. ({\bf right}) The lowest $\deltav$ transfer from the distribution at left at $88\mps$. The Moon's orbit is shown as the blue ellipse around the green Earth. The thin grey path is the orbit of the TCO starting from its capture point (green triangle) to its escape point (red square). The blue circle on the TCO orbit marks where the TCO is when the spacecraft departs from its halo orbit, and the yellow star represents the  rendezvous location. The magenta path is the spacecraft's trajectory and its three burn maneuvers are marked as yellow dots (including the final rendezvous burn).}
    
    \label{fig.MinimoonMissionOpportunities}

\end{center}
\end{figure}

The second technique to overcome the limitation of the short-duration minimoon captures is to maintain a spacecraft in a `hibernating' orbit awaiting the arrival and discovery of a suitably interesting minimoon. This idea may seem untenable at this time but will become practical once LSST begins discovering many minimoons per month (\S\ref{s.MinimoonCurrent+FutureStatus}).  Unlike distant asteroids, minimoon orbits can be rapidly and accurately determined (\fig{fig.MinimoonOrbitlUncertaintyEvolution}) to enable this opportunity and could even allow for multiple minimoon missions with the same spacecraft.

With this technique in mind, minimoon rendezvous missions have been studied using indirect (minimization) methods within the circular restricted four-body problem (CR4BP; Sun, Earth, Moon, spacecraft) with the Sun acting as a perturbation on the Earth-Moon-spacecraft CR3BP \citep{Chyba2016,Brelsford2016}.  Using a random sample of 3,000 TCOs from \citet{Granvik2012} they showed that rendezvous trajectories could be designed for all of them with a median $\deltav$ just under $680\mps$ (\fig{fig.MinimoonMissionOpportunities}) with most of the transfer durations, the time from EM \Ltwo\ departure to minimoon rendezvous, requiring less than three months.  The mean $\deltav=725\mps$ is about 7\% higher than the median due to a tail of high $\deltav$ transfers but the minimum $\deltav$ is only $88\mps$ with a transfer time of $41\days$ (\fig{fig.MinimoonMissionOpportunities}).  
Even more intriguing, in a future where the LSST is discovering all the large minimoon captures on a regular basis, we can envision multiple successive minimoon rendezvous missions with transfers directly between the minimoons.  As a first step to modeling this possibility \citet{Chyba2016} examined round trip mission opportunities for TCO \RH\ because, in a worst case scenario, multiple minimoon missions could simply be back-to-back missions from the EM \Ltwo\ hibernating halo orbit (they assumed a $z$-excursion of 5,000$\km$ in the halo orbit). The round trip is composed of a transfer to bring the spacecraft to \RH, followed by a rendezvous phase where the spacecraft travels with the asteroid, and finally a return transfer back to the hibernating orbit. The lowest round-trip $\deltav$ required only $901\mps$ with a total duration of $630\days$ ($173\days$ for the approach and $240\days$ for the return) including $217\days$ at the asteroid.

\section{Conclusions}
\label{s.Conclusions}

Earth's minimoons will provide an opportunity for low-$\deltav$ scientific exploration and commercial exploitation of small asteroids where most of the effort of bringing the objects to Earth has been accomplished by their slow dynamical evolution from the main belt.  While naturally produced minimoons will be too small for commercially profitable enterprises they will be extremely useful for testing techniques in a cis-lunar environment before moving operations into distant heliocentric space.  There are also opportunities of artificially enhancing the minimoon population by selectively maneuvering scientifically or commercially small asteroids onto geocentric capture trajectories from their heliocentric orbits.  

The challenge in minimoon studies or capture is discovering them.  Naturally produced minimoons are small, with the largest in the steady state population being perhaps only $1\meter$ in diameter.  Enhancing the minimoon capture rate requires detecting decameter-scale asteroids long before they enter Earth's Hill sphere.

The Large Synoptic Survey Telescope will be capable of detecting the largest natural minimoons and will also detect a substantial number of NEOs that could be artificially induced into becoming minimoons but the real future for mining asteroids awaits an affordable space-based detection system.  Once those assets are in place they will unlock the exploration of the solar system with minimoons being the first stepping stones.

\section*{Conflict of Interest Statement}

The authors declare that the research was conducted in the absence of any commercial or financial relationships that could be construed as a potential conflict of interest.


\section*{Author Contributions}


The author order is alphabetical after the first author.  RJ coordinated the entire effort and wrote the majority of the text.  BTB provided insight into future prospects for minimoon discovery and assisted writing the text on the area-to-mass ratio.  WFB provided text on the dynamical aspects of minimoons including the Yarkovsky and lunar-origin material.  MC provided material on simulating and minimizing minimoon mission $\deltav$.  GF provided details from his paper on the minimoon population and opportunities for minimoon discovery with LSST.  MG provided details from his paper on the minimoon population and perspective on the science and mission opportunities.  LJ provided her expertise and results of LSST minimoon discovery capabilities.  HU contributed the second largest amount to the paper with significant text regarding minimoon dynamics.

\section*{Funding}

WFB’s participation was supported by NASA’s SSERVI program “Institute for the Science of Exploration Targets (ISET)” through institute grant number NNA14AB03A.  RJ was partly supported by WFB.  MC was partially supported by award \# 359510 from the Simons Foundation.  MG is partially funded by grant \#299543 from the Academy of Finland.  HU wishes to acknowledge funding from grant ESP2017-87271-P (MINECO/AEI/FEDER, UE).  GF was supported in part by a grant from the Emil Aaltonen foundation.

\section*{Acknowledgments}

RJ thanks Giovanni Valsecchi (INAF \& IAPS, Rome, Italy) and Marco Micheli (ESA SSA-NEO Coordination Centre, Frascati, Italy) for their insight and support in both dynamical and observational aspects of minimoons.  We thank three reviewers and the editors for  helpful suggestions to improve the review in many ways.


\bibliographystyle{frontiersinSCNS_ENG_HUMS} 
\bibliography{references}

\begin{thebibliography}{102}
\providecommand{\natexlab}[1]{#1}
\expandafter\ifx\csname urlstyle\endcsname\relax
  \providecommand{\doi}[1]{doi:\discretionary{}{}{}#1}\else
  \providecommand{\doi}{doi:\discretionary{}{}{}\begingroup
  \urlstyle{rm}\Url}\fi
\providecommand{\selectlanguage}[1]{\relax}
\providecommand{\bibAnnoteFile}[1]{%
  \IfFileExists{#1}{\begin{quotation}\noindent\textsc{Key:} #1\\
  \textsc{Annotation:}\ \input{#1}\end{quotation}}{}}
\providecommand{\bibAnnote}[2]{%
  \begin{quotation}\noindent\textsc{Key:} #1\\
  \textsc{Annotation:}\ #2\end{quotation}}

\bibitem[{Astakhov et~al.(2003)Astakhov, Burbanks, Wiggins, and
  Farrelly}]{Astakhov2003}
Astakhov, S.~A., Burbanks, A.~D., Wiggins, S., and Farrelly, D. (2003).
\newblock {Chaos-assisted capture of irregular moons}.
\newblock \emph{Nature} 423, 264–267.
\newblock \doi{10.1038/nature01622}
\bibAnnoteFile{Astakhov2003}

\bibitem[{Bao et~al.(2015)Bao, Yang, Barsbold, and Baoyin}]{Bao2015}
Bao, C., Yang, H., Barsbold, B., and Baoyin, H. (2015).
\newblock Capturing near-earth asteroids into bounded earth orbits using
  gravity assist.
\newblock \emph{Astrophysics and Space Science} 360, 61.
\newblock \doi{10.1007/s10509-015-2581-3}
\bibAnnoteFile{Bao2015}

\bibitem[{Baoyin et~al.(2010)Baoyin, Chen, and Li}]{Baoyin2010}
Baoyin, H.-X., Chen, Y., and Li, J.-F. (2010).
\newblock Capturing near earth objects.
\newblock \emph{Research in Astronomy and Astrophysics} 10, 587
\bibAnnoteFile{Baoyin2010}

\bibitem[{{Benner} et~al.(2015){Benner}, {Busch}, {Giorgini}, {Taylor}, and
  {Margot}}]{Benner2015-AsteroidsIV}
{Benner}, L.~A.~M., {Busch}, M.~W., {Giorgini}, J.~D., {Taylor}, P.~A., and
  {Margot}, J.-L. (2015).
\newblock \emph{{Radar Observations of Near-Earth and Main-Belt Asteroids}}.
\newblock 165--182.
\newblock \doi{10.2458/azu_uapress_9780816532131-ch009}
\bibAnnoteFile{Benner2015-AsteroidsIV}

\bibitem[{Beutler et~al.(2006)Beutler, Mervart, and Verdun}]{Beutler2006}
Beutler, G., Mervart, L., and Verdun, A. (2006).
\newblock \emph{Methods of Celestial Mechanics: Volume II: Application to
  Planetary System, Geodynamics and Satellite Geodesy}.
\newblock Astronomy and Astrophysics Library (Springer Berlin Heidelberg)
\bibAnnoteFile{Beutler2006}

\bibitem[{{Bolin} et~al.(2014){Bolin}, {Jedicke}, {Granvik}, {Brown}, {Howell},
  {Nolan} et~al.}]{Bolin2014}
{Bolin}, B., {Jedicke}, R., {Granvik}, M., {Brown}, P., {Howell}, E., {Nolan},
  M.~C., et~al. (2014).
\newblock {Detecting Earth's temporarily-captured natural
  satellites-Minimoons}.
\newblock \emph{\icarus} 241, 280--297.
\newblock \doi{10.1016/j.icarus.2014.05.026}
\bibAnnoteFile{Bolin2014}

\bibitem[{{Bottke} et~al.(2002){Bottke}, {Morbidelli}, {Jedicke}, {Petit},
  {Levison}, {Michel} et~al.}]{Bottke2002}
{Bottke}, W.~F., {Morbidelli}, A., {Jedicke}, R., {Petit}, J.-M., {Levison},
  H.~F., {Michel}, P., et~al. (2002).
\newblock {Debiased Orbital and Absolute Magnitude Distribution of the
  Near-Earth Objects}.
\newblock \emph{\icarus} 156, 399--433.
\newblock \doi{10.1006/icar.2001.6788}
\bibAnnoteFile{Bottke2002}

\bibitem[{{Bottke} et~al.(2006){Bottke}, {Vokrouhlick{\'y}}, {Rubincam}, and
  {Nesvorn{\'y}}}]{Bottke2006-Yarko+YORP-Review}
{Bottke}, W.~F., Jr., {Vokrouhlick{\'y}}, D., {Rubincam}, D.~P., and
  {Nesvorn{\'y}}, D. (2006).
\newblock {The Yarkovsky and Yorp Effects: Implications for Asteroid Dynamics}.
\newblock \emph{Annual Review of Earth and Planetary Sciences} 34, 157--191.
\newblock \doi{10.1146/annurev.earth.34.031405.125154}
\bibAnnoteFile{Bottke2006-Yarko+YORP-Review}

\bibitem[{{Brelsford} et~al.(2016){Brelsford}, {Chyba}, {Haberkorn}, and
  {Patterson}}]{Brelsford2016}
{Brelsford}, S., {Chyba}, M., {Haberkorn}, T., and {Patterson}, G. (2016).
\newblock {Rendezvous missions to temporarily captured near Earth asteroids}.
\newblock \emph{\planss} 123, 4--15.
\newblock \doi{10.1016/j.pss.2015.12.013}
\bibAnnoteFile{Brelsford2016}

\bibitem[{Brophy et~al.(2012)Brophy, Culick, Friedman, Allen, Baughman,
  Bellerose et~al.}]{Brophy2012}
Brophy, J., Culick, F., Friedman, L., Allen, C., Baughman, D., Bellerose, J.,
  et~al. (2012).
\newblock \emph{Asteroid Retrieval Feasibility Study}.
\newblock Tech. rep., Keck Institute for Space Studies, California Institute of
  Technology, Jet Propulsion Laboratory
\bibAnnoteFile{Brophy2012}

\bibitem[{{Brophy} and {Muirhead}(2013)}]{Brophy2013}
{Brophy}, J.~R. and {Muirhead}, B. (2013).
\newblock \emph{AsteroidNear-Earth Asteroid Retrieval Mission (ARM) study}.
\newblock Tech. rep., Keck Institute for Space Studies, California Institute of
  Technology, Jet Propulsion Laboratory
\bibAnnoteFile{Brophy2013}

\bibitem[{{Brown} et~al.(2013){Brown}, {Assink}, {Astiz}, {Blaauw}, {Boslough},
  {Borovi{\v c}ka} et~al.}]{Brown2013}
{Brown}, P.~G., {Assink}, J.~D., {Astiz}, L., {Blaauw}, R., {Boslough}, M.~B.,
  {Borovi{\v c}ka}, J., et~al. (2013).
\newblock {A 500-kiloton airburst over Chelyabinsk and an enhanced hazard from
  small impactors}.
\newblock \emph{\nat} 503, 238--241.
\newblock \doi{10.1038/nature12741}
\bibAnnoteFile{Brown2013}

\bibitem[{Brown et~al.(2010)Brown, Weryk, Kohut, Edwards, and
  Krzemenski}]{Brown2010}
Brown, P.~G., Weryk, R., Kohut, S., Edwards, W.~N., and Krzemenski, Z. (2010).
\newblock Development of an all-sky video meteor network in southern ontario,
  canada: The asgard system.
\newblock \emph{J. IMO} 38
\bibAnnoteFile{Brown2010}

\bibitem[{{Carry}(2012)}]{Carry2012}
{Carry}, B. (2012).
\newblock {Density of asteroids}.
\newblock \emph{\planss} 73, 98--118.
\newblock \doi{10.1016/j.pss.2012.03.009}
\bibAnnoteFile{Carry2012}

\bibitem[{Carusi and Valsecchi(1981)}]{Carusi1981}
Carusi, A. and Valsecchi, G.~B. (1981).
\newblock Temporary satellite captures of comets by jupiter.
\newblock \emph{Astronomy and Astrophysics} 94, 226--228
\bibAnnoteFile{Carusi1981}

\bibitem[{{Chant}(1913{\natexlab{a}})}]{Chant1913a}
{Chant}, C.~A. (1913{\natexlab{a}}).
\newblock {An Extraordinary Meteoric Display}.
\newblock \emph{\jrasc} 7, 145
\bibAnnoteFile{Chant1913a}

\bibitem[{{Chant}(1913{\natexlab{b}})}]{Chant1913b}
{Chant}, C.~A. (1913{\natexlab{b}}).
\newblock {Further Information Regarding the Meteoric Display of February 9
  1913}.
\newblock \emph{\jrasc} 7, 438
\bibAnnoteFile{Chant1913b}

\bibitem[{{Chen}(2016)}]{Chen2016a}
{Chen}, H. (2016).
\newblock {Analysis and Design of Asteroid Retrieval Missions Using Luni-Solar
  Gravity Assists}.
\newblock In \emph{41st COSPAR Scientific Assembly}. vol.~41 of \emph{COSPAR
  Meeting}
\bibAnnoteFile{Chen2016a}

\bibitem[{{Chen} et~al.(2016){Chen}, {Lin}, {Holman}, {Payne}, {Fraser},
  {Lacerda} et~al.}]{Chen2016-RetrogradeTNO}
{Chen}, Y.-T., {Lin}, H.~W., {Holman}, M.~J., {Payne}, M.~J., {Fraser}, W.~C.,
  {Lacerda}, P., et~al. (2016).
\newblock {Discovery of a New Retrograde Trans-Neptunian Object: Hint of a
  Common Orbital Plane for Low Semimajor Axis, High-inclination TNOs and
  Centaurs}.
\newblock \emph{\apjl} 827, L24.
\newblock \doi{10.3847/2041-8205/827/2/L24}
\bibAnnoteFile{Chen2016-RetrogradeTNO}

\bibitem[{{Chodas}(2016)}]{Chodas2016DPS-HO3}
{Chodas}, P. (2016).
\newblock {The Orbit and Future Motion of Earth Quasi-Satellite 2016 HO3}.
\newblock In \emph{AAS/Division for Planetary Sciences Meeting Abstracts \#48}.
  vol.~48 of \emph{AAS/Division for Planetary Sciences Meeting Abstracts},
  311.04
\bibAnnoteFile{Chodas2016DPS-HO3}

\bibitem[{{Chodas} and {Yeomans}(1996)}]{Chodas1996-SL9}
{Chodas}, P.~W. and {Yeomans}, D.~K. (1996).
\newblock {The orbital motion and impact circumstances of Comet Shoemaker-Levy
  9}.
\newblock In \emph{IAU Colloq. 156: The Collision of Comet Shoemaker-Levy 9 and
  Jupiter}, eds. K.~S. {Noll}, H.~A. {Weaver}, and P.~D. {Feldman}. 1--30
\bibAnnoteFile{Chodas1996-SL9}

\bibitem[{{Chyba} et~al.(2016){Chyba}, {Haberkorn}, and
  {Patterson}}]{Chyba2016}
{Chyba}, M., {Haberkorn}, T., and {Patterson}, G. (2016).
\newblock {Rendezvous Missions to Temporarily-Captured Near Earth Asteroids}.
\newblock \emph{Planetary and Space Science} 123, 4--15
\bibAnnoteFile{Chyba2016}

\bibitem[{{Clark} et~al.(2016){Clark}, {Spurn{\'y}}, {Wiegert}, {Brown},
  {Borovi{\v c}ka}, {Tagliaferri} et~al.}]{Clark2016}
{Clark}, D.~L., {Spurn{\'y}}, P., {Wiegert}, P., {Brown}, P., {Borovi{\v c}ka},
  J., {Tagliaferri}, E., et~al. (2016).
\newblock {Impact Detections of Temporarily Captured Natural Satellites}.
\newblock \emph{\aj} 151, 135.
\newblock \doi{10.3847/0004-6256/151/6/135}
\bibAnnoteFile{Clark2016}

\bibitem[{{Connors} et~al.(2011){Connors}, {Wiegert}, and
  {Veillet}}]{Connors2011-EarthTrojan}
{Connors}, M., {Wiegert}, P., and {Veillet}, C. (2011).
\newblock {Earth's Trojan asteroid}.
\newblock \emph{\nat} 475, 481--483.
\newblock \doi{10.1038/nature10233}
\bibAnnoteFile{Connors2011-EarthTrojan}

\bibitem[{{Cyr} et~al.(2000){Cyr}, {Mesarch}, {Maldonado}, {Folta}, {Harper},
  {Davila} et~al.}]{Cyr2000-SpaceWeatherDiamond}
{Cyr}, O.~C.~S., {Mesarch}, M.~A., {Maldonado}, H.~M., {Folta}, D.~C.,
  {Harper}, A.~D., {Davila}, J.~M., et~al. (2000).
\newblock {Space Weather Diamond: a four spacecraft monitoring system}.
\newblock \emph{Journal of Atmospheric and Solar-Terrestrial Physics} 62,
  1251--1255.
\newblock \doi{10.1016/S1364-6826(00)00069-9}
\bibAnnoteFile{Cyr2000-SpaceWeatherDiamond}

\bibitem[{{de la Fuente Marcos} and {de la Fuente
  Marcos}(2013)}]{FuenteMarcos2013}
{de la Fuente Marcos}, C. and {de la Fuente Marcos}, R. (2013).
\newblock {A resonant family of dynamically cold small bodies in the near-Earth
  asteroid belt}.
\newblock \emph{\mnras} 434, L1--L5.
\newblock \doi{10.1093/mnrasl/slt062}
\bibAnnoteFile{FuenteMarcos2013}

\bibitem[{{de la Fuente Marcos} and {de la Fuente
  Marcos}(2016)}]{FuenteMarcos2016-HO3}
{de la Fuente Marcos}, C. and {de la Fuente Marcos}, R. (2016).
\newblock {Asteroid (469219) 2016 HO$_{3}$, the smallest and closest Earth
  quasi-satellite}.
\newblock \emph{\mnras} 462, 3441--3456.
\newblock \doi{10.1093/mnras/stw1972}
\bibAnnoteFile{FuenteMarcos2016-HO3}

\bibitem[{{Denneau} et~al.(2013){Denneau}, {Jedicke}, {Grav}, {Granvik},
  {Kubica}, {Milani} et~al.}]{Denneau2013}
{Denneau}, L., {Jedicke}, R., {Grav}, T., {Granvik}, M., {Kubica}, J.,
  {Milani}, A., et~al. (2013).
\newblock {The Pan-STARRS Moving Object Processing System}.
\newblock \emph{\pasp} 125, 357--395.
\newblock \doi{10.1086/670337}
\bibAnnoteFile{Denneau2013}

\bibitem[{{Denning}(1916)}]{Denning1916}
{Denning}, W.~F. (1916).
\newblock {The Remarkable Meteors of February 9, 1913}.
\newblock \emph{\nat} 97, 181.
\newblock \doi{10.1038/097181b0}
\bibAnnoteFile{Denning1916}

\bibitem[{{Elvis} et~al.(2011){Elvis}, {McDowell}, {Hoffman}, and
  {Binzel}}]{Elvis2011}
{Elvis}, M., {McDowell}, J., {Hoffman}, J.~A., and {Binzel}, R.~P. (2011).
\newblock {Ultra-low delta-v objects and the human exploration of asteroids}.
\newblock \emph{\planss} 59, 1408--1412.
\newblock \doi{10.1016/j.pss.2011.05.006}
\bibAnnoteFile{Elvis2011}

\bibitem[{{Eugster} et~al.(2006){Eugster}, {Herzog}, {Marti}, and
  {Caffee}}]{Eugster2006-MESSII}
{Eugster}, O., {Herzog}, G.~F., {Marti}, K., and {Caffee}, M.~W. (2006).
\newblock \emph{{Irradiation Records, Cosmic-Ray Exposure Ages, and Transfer
  Times of Meteorites}}.
\newblock 829--851
\bibAnnoteFile{Eugster2006-MESSII}

\bibitem[{{Farnocchia} et~al.(2016){Farnocchia}, {Chesley}, and
  {Chamberlin}}]{Farnocchia2016-DPS-Scout}
{Farnocchia}, D., {Chesley}, S.~R., and {Chamberlin}, A.~B. (2016).
\newblock {Scout: orbit analysis and hazard assessment for NEOCP objects}.
\newblock In \emph{AAS/Division for Planetary Sciences Meeting Abstracts \#48}.
  vol.~48 of \emph{AAS/Division for Planetary Sciences Meeting Abstracts},
  305.03
\bibAnnoteFile{Farnocchia2016-DPS-Scout}

\bibitem[{{Farnocchia} et~al.(2017){Farnocchia}, {Tholen}, {Micheli}, {Ryan},
  {Rivera-Valentin}, {Taylor} et~al.}]{Farnocchia2017-DPS-2015TC25}
{Farnocchia}, D., {Tholen}, D.~J., {Micheli}, M., {Ryan}, W.,
  {Rivera-Valentin}, E.~G., {Taylor}, P.~A., et~al. (2017).
\newblock {Mass estimate and close approaches of near-Earth asteroid 2015
  TC25}.
\newblock In \emph{AAS/Division for Planetary Sciences Meeting Abstracts \#49}.
  vol.~49 of \emph{AAS/Division for Planetary Sciences Meeting Abstracts},
  100.09
\bibAnnoteFile{Farnocchia2017-DPS-2015TC25}

\bibitem[{{Fedorets} et~al.(2017){Fedorets}, {Granvik}, and
  {Jedicke}}]{Fedorets2017}
{Fedorets}, G., {Granvik}, M., and {Jedicke}, R. (2017).
\newblock {Orbit and size distributions for asteroids temporarily captured by
  the Earth-Moon system}.
\newblock \emph{\icarus} 285, 83--94.
\newblock \doi{10.1016/j.icarus.2016.12.022}
\bibAnnoteFile{Fedorets2017}

\bibitem[{{Fedorets} et~al.(2015){Fedorets}, {Granvik}, {Jones}, and
  {Jedicke}}]{Fedorets2015-LSST}
{Fedorets}, G., {Granvik}, M., {Jones}, L., and {Jedicke}, R. (2015).
\newblock {Discovering asteroids temporarily captured by the Earth with LSST}.
\newblock \emph{IAU General Assembly} 22, 2257052
\bibAnnoteFile{Fedorets2015-LSST}

\bibitem[{{Garc{\'{\i}}a Y{\'a}rnoz} et~al.(2013{\natexlab{a}}){Garc{\'{\i}}a
  Y{\'a}rnoz}, {Sanchez}, and {McInnes}}]{Garcia2013}
{Garc{\'{\i}}a Y{\'a}rnoz}, D., {Sanchez}, J.~P., and {McInnes}, C.~R.
  (2013{\natexlab{a}}).
\newblock {Easily retrievable objects among the NEO population}.
\newblock \emph{Celestial Mechanics and Dynamical Astronomy} 116, 367--388.
\newblock \doi{10.1007/s10569-013-9495-6}
\bibAnnoteFile{Garcia2013}

\bibitem[{{Garc{\'{\i}}a Y{\'a}rnoz} et~al.(2013{\natexlab{b}}){Garc{\'{\i}}a
  Y{\'a}rnoz}, {Sanchez}, and {McInnes}}]{Garcia2013a}
{Garc{\'{\i}}a Y{\'a}rnoz}, D., {Sanchez}, J.-P., and {McInnes}, C.~R.
  (2013{\natexlab{b}}).
\newblock \emph{{Opportunities for Asteroid Retrieval Missions}}.
\newblock 479--505.
\newblock \doi{10.1007/978-3-642-39244-3_21}
\bibAnnoteFile{Garcia2013a}

\bibitem[{Gong and Li(2015)}]{Gong2015}
Gong, S. and Li, J. (2015).
\newblock Asteroid capture using lunar flyby.
\newblock \emph{Advances in Space Research} 56, 848 -- 858.
\newblock \doi{10.1016/j.asr.2015.05.020}
\bibAnnoteFile{Gong2015}

\bibitem[{{Granvik} et~al.(2013){Granvik}, {Jedicke}, {Bolin}, {Chyba}, and
  {Patterson}}]{Granvik2013}
{Granvik}, M., {Jedicke}, R., {Bolin}, B., {Chyba}, M., and {Patterson}, G.
  (2013).
\newblock \emph{{Earth's Temporarily-Captured Natural Satellites - The First
  Step towards Utilization of Asteroid Resources}}.
\newblock 151--167.
\newblock \doi{10.1007/978-3-642-39244-3_6}
\bibAnnoteFile{Granvik2013}

\bibitem[{{Granvik} et~al.(2016){Granvik}, {Morbidelli}, {Jedicke}, {Bolin},
  {Bottke}, {Beshore} et~al.}]{Granvik2016a-NEOModel-Nature}
{Granvik}, M., {Morbidelli}, A., {Jedicke}, R., {Bolin}, B., {Bottke}, W.~F.,
  {Beshore}, E., et~al. (2016).
\newblock {Super-catastrophic disruption of asteroids at small perihelion
  distances}.
\newblock \emph{\nat} 530, 303--306.
\newblock \doi{10.1038/nature16934}
\bibAnnoteFile{Granvik2016a-NEOModel-Nature}

\bibitem[{{Granvik} et~al.(2017){Granvik}, {Morbidelli}, {Vokrouhlick{\'y}},
  {Bottke}, {Nesvorn{\'y}}, and {Jedicke}}]{Granvik2017-escapeFromMainBelt}
{Granvik}, M., {Morbidelli}, A., {Vokrouhlick{\'y}}, D., {Bottke}, W.~F.,
  {Nesvorn{\'y}}, D., and {Jedicke}, R. (2017).
\newblock {Escape of asteroids from the main belt}.
\newblock \emph{\aap} 598, A52.
\newblock \doi{10.1051/0004-6361/201629252}
\bibAnnoteFile{Granvik2017-escapeFromMainBelt}

\bibitem[{{Granvik} et~al.(2012){Granvik}, {Vaubaillon}, and
  {Jedicke}}]{Granvik2012}
{Granvik}, M., {Vaubaillon}, J., and {Jedicke}, R. (2012).
\newblock {The population of natural Earth satellites}.
\newblock \emph{\icarus} 218, 262--277.
\newblock \doi{10.1016/j.icarus.2011.12.003}
\bibAnnoteFile{Granvik2012}

\bibitem[{{Graps} et~al.(2016){Graps}, {Blondel}, {Bonin}, {Britt}, {Centuori},
  {Delbo} et~al.}]{Graps2016}
{Graps}, A.~L., {Blondel}, P., {Bonin}, G., {Britt}, D., {Centuori}, S.,
  {Delbo}, M., et~al. (2016).
\newblock {ASIME 2016 White Paper: In-Space Utilisation of Asteroids: ``Answers
  to Questions from the Asteroid Miners''}.
\newblock \emph{ArXiv e-prints}
\bibAnnoteFile{Graps2016}

\bibitem[{{Gural} et~al.(2005){Gural}, {Larsen}, and
  {Gleason}}]{Gural2005-MatcherFilter-Spacewatch}
{Gural}, P.~S., {Larsen}, J.~A., and {Gleason}, A.~E. (2005).
\newblock {Matched Filter Processing for Asteroid Detection}.
\newblock \emph{\aj} 130, 1951--1960.
\newblock \doi{10.1086/444415}
\bibAnnoteFile{Gural2005-MatcherFilter-Spacewatch}

\bibitem[{{Harris} et~al.(2016){Harris}, {Morbidelli}, and
  {Granvik}}]{Harris2016DPS-LifeNearZero}
{Harris}, A.~W., {Morbidelli}, A., and {Granvik}, M. (2016).
\newblock {Life and Death Near Zero: The distribution and evolution of NEA
  orbits of near-zero MOID, (e, i), and q}.
\newblock In \emph{AAS/Division for Planetary Sciences Meeting Abstracts \#48}.
  vol.~48 of \emph{AAS/Division for Planetary Sciences Meeting Abstracts},
  305.05
\bibAnnoteFile{Harris2016DPS-LifeNearZero}

\bibitem[{Hasnain et~al.(2012)Hasnain, Lamb, and Ross}]{Hasnain2012}
Hasnain, Z., Lamb, C.~A., and Ross, S.~D. (2012).
\newblock Capturing near-earth asteroids around earth.
\newblock \emph{Acta Astronautica} 81, 523--531.
\newblock \doi{https://doi.org/10.1016/j.actaastro.2012.07.029}
\bibAnnoteFile{Hasnain2012}

\bibitem[{{Heinze} et~al.(2015){Heinze}, {Metchev}, and
  {Trollo}}]{Heinze2015-DigitalTracking}
{Heinze}, A.~N., {Metchev}, S., and {Trollo}, J. (2015).
\newblock {Digital Tracking Observations Can Discover Asteroids 10 Times
  Fainter Than Conventional Searches}.
\newblock \emph{\aj} 150, 125.
\newblock \doi{10.1088/0004-6256/150/4/125}
\bibAnnoteFile{Heinze2015-DigitalTracking}

\bibitem[{Heppenheimer and Porco(1977)}]{Heppenheimer1977}
Heppenheimer, T.~A. and Porco, C. (1977).
\newblock New contributions to the problem of capture.
\newblock \emph{Icarus} 30, 385--401
\bibAnnoteFile{Heppenheimer1977}

\bibitem[{{Hills} and {Goda}(1997)}]{Hills1997-GrazingMeteoroids}
{Hills}, J.~G. and {Goda}, M.~P. (1997).
\newblock {Meteoroids captured into Earth orbit by grazing atmospheric
  encounters}.
\newblock \emph{\planss} 45, 595--602.
\newblock \doi{10.1016/S0032-0633(97)00039-1}
\bibAnnoteFile{Hills1997-GrazingMeteoroids}

\bibitem[{{Hou} et~al.(2015){Hou}, {Xin}, {Scheeres}, and
  {Wang}}]{Hou2015-EM-Trojans}
{Hou}, X.~Y., {Xin}, X., {Scheeres}, D.~J., and {Wang}, J. (2015).
\newblock {Stable motions around triangular libration points in the real
  Earth-Moon system}.
\newblock \emph{\mnras} 454, 4172--4181.
\newblock \doi{10.1093/mnras/stv2216}
\bibAnnoteFile{Hou2015-EM-Trojans}

\bibitem[{Huang and Innanen(1983)}]{Huang1983}
Huang, T.-Y. and Innanen, K.~A. (1983).
\newblock The gravitational espace/capture of planetary satellite.
\newblock \emph{The Astronomical Journal} 88, 1537--1548
\bibAnnoteFile{Huang1983}

\bibitem[{Hunt et~al.(2004)Hunt, Oppenheim, Close, Brown, McKeen, and
  Minardi}]{Hunt2004-MeteorVelocity}
Hunt, S., Oppenheim, M., Close, S., Brown, P., McKeen, F., and Minardi, M.
  (2004).
\newblock Determination of the meteoroid velocity distribution at the earth
  using high-gain radar.
\newblock \emph{Icarus} 168.
\newblock \doi{10.1016/j.icarus.2003.08.006}
\bibAnnoteFile{Hunt2004-MeteorVelocity}

\bibitem[{Hutcheon(2013)}]{Olson2013}
[Dataset] Hutcheon, O.~. (2013).
\newblock The great meteor procession of 1913
\bibAnnoteFile{Olson2013}

\bibitem[{{Ivezic} et~al.(2008){Ivezic}, {Axelrod}, {Brandt}, {Burke},
  {Claver}, {Connolly} et~al.}]{Ivezic2008-LSST}
{Ivezic}, Z., {Axelrod}, T., {Brandt}, W.~N., {Burke}, D.~L., {Claver}, C.~F.,
  {Connolly}, A., et~al. (2008).
\newblock {Large Synoptic Survey Telescope: From Science Drivers To Reference
  Design}.
\newblock \emph{Serbian Astronomical Journal} 176, 1--13.
\newblock \doi{10.2298/SAJ0876001I}
\bibAnnoteFile{Ivezic2008-LSST}

\bibitem[{{Jedicke} et~al.(2017{\natexlab{a}}){Jedicke}, {Boe}, {Bolin},
  {Bottke}, {Chyba}, {Denneau} et~al.}]{Jedicke2017-DPS-minimoonsJ}
{Jedicke}, R., {Boe}, B., {Bolin}, B.~T., {Bottke}, W., {Chyba}, M., {Denneau},
  L., et~al. (2017{\natexlab{a}}).
\newblock {Minimoon Survey with Subaru Hyper Suprime-Cam}.
\newblock In \emph{AAS/Division for Planetary Sciences Meeting Abstracts \#49}.
  vol.~49 of \emph{AAS/Division for Planetary Sciences Meeting Abstracts},
  112.01
\bibAnnoteFile{Jedicke2017-DPS-minimoonsJ}

\bibitem[{{Jedicke} et~al.(2018){Jedicke}, {Sercel}, {Gillis-Davis}, {Morenz},
  and {Gertsch}}]{Jedicke2018}
{Jedicke}, R., {Sercel}, J., {Gillis-Davis}, J., {Morenz}, K.~J., and
  {Gertsch}, L. (2018).
\newblock {Availability and delta-v requirements for delivering water extracted
  from near-Earth objects to cis-lunar space}.
\newblock \emph{\planss} 159, 28--42.
\newblock \doi{10.1016/j.pss.2018.04.005}
\bibAnnoteFile{Jedicke2018}

\bibitem[{{Jedicke} et~al.(2017{\natexlab{b}}){Jedicke}, {Sercel}, {Morenz},
  and {Gertsch}}]{Jedicke2017DDA}
{Jedicke}, R., {Sercel}, J., {Morenz}, K.~J., and {Gertsch}, L.~S.
  (2017{\natexlab{b}}).
\newblock {Broken Plane Delta-v Calculation For Rapid Assessment of Synthetic
  Asteroid Targets for In-situ Resource Utilization}.
\newblock In \emph{AAS/Division of Dynamical Astronomy Meeting \#48}. vol.~48
  of \emph{AAS/Division of Dynamical Astronomy Meeting}, 205.03
\bibAnnoteFile{Jedicke2017DDA}

\bibitem[{{Jenniskens} et~al.(2011){Jenniskens}, {Gural}, {Dynneson},
  {Grigsby}, {Newman}, {Borden} et~al.}]{Jenniskens2011}
{Jenniskens}, P., {Gural}, P.~S., {Dynneson}, L., {Grigsby}, B.~J., {Newman},
  K.~E., {Borden}, M., et~al. (2011).
\newblock {CAMS: Cameras for Allsky Meteor Surveillance to establish minor
  meteor showers}.
\newblock \emph{\icarus} 216, 40--61.
\newblock \doi{10.1016/j.icarus.2011.08.012}
\bibAnnoteFile{Jenniskens2011}

\bibitem[{{Jorgensen} et~al.(2003){Jorgensen}, {Rivkin}, {Binzel}, {Whitely},
  {Hergenrother}, {Chodas} et~al.}]{Jorgensen2003-J002E3}
{Jorgensen}, K., {Rivkin}, A., {Binzel}, R., {Whitely}, R., {Hergenrother}, C.,
  {Chodas}, P., et~al. (2003).
\newblock {Observations of J002E3: Possible Discovery of an Apollo Rocket
  Body}.
\newblock In \emph{AAS/Division for Planetary Sciences Meeting Abstracts \#35}.
  vol.~35 of \emph{Bulletin of the American Astronomical Society}, 981
\bibAnnoteFile{Jorgensen2003-J002E3}

\bibitem[{{Kary} and {Dones}(1996)}]{Kary1996-SPC-CaptureStatistics}
{Kary}, D.~M. and {Dones}, L. (1996).
\newblock {Capture Statistics of Short-Period Comets: Implications for Comet
  D/Shoemaker-Levy 9}.
\newblock \emph{\icarus} 121, 207--224.
\newblock \doi{10.1006/icar.1996.0082}
\bibAnnoteFile{Kary1996-SPC-CaptureStatistics}

\bibitem[{Koon et~al.(2001)Koon, Lo, Marsden, and Ross}]{Koon2001}
Koon, W.~S., Lo, M.~W., Marsden, J.~E., and Ross, S.~D. (2001).
\newblock Resonance and capture of jupiter comets.
\newblock \emph{Celestial Mechanics and Dynamical Astronomy} 81, 27--38.
\newblock \doi{10.1023/A:1013398801813}
\bibAnnoteFile{Koon2001}

\bibitem[{{Kwiatkowski} et~al.(2009){Kwiatkowski}, {Kryszczy{\'n}ska},
  {Poli{\'n}ska}, {Buckley}, {O'Donoghue}, {Charles} et~al.}]{Kwiatkowski2009}
{Kwiatkowski}, T., {Kryszczy{\'n}ska}, A., {Poli{\'n}ska}, M., {Buckley},
  D.~A.~H., {O'Donoghue}, D., {Charles}, P.~A., et~al. (2009).
\newblock {Photometry of 2006 RH$\{$120$\}$: an asteroid temporary captured
  into a geocentric orbit}.
\newblock \emph{\aap} 495, 967--974.
\newblock \doi{10.1051/0004-6361:200810965}
\bibAnnoteFile{Kwiatkowski2009}

\bibitem[{{Larson} et~al.(1998){Larson}, {Brownlee}, {Hergenrother}, and
  {Spahr}}]{Larson1998}
{Larson}, S., {Brownlee}, J., {Hergenrother}, C., and {Spahr}, T. (1998).
\newblock {The Catalina Sky Survey for NEOs}.
\newblock In \emph{Bulletin of the American Astronomical Society}. vol.~30 of
  \emph{Bulletin of the American Astronomical Society}, 1037
\bibAnnoteFile{Larson1998}

\bibitem[{{Mainzer} et~al.(2011){Mainzer}, {Bauer}, {Grav}, {Masiero}, {Cutri},
  {Dailey} et~al.}]{Mainzer2011a}
{Mainzer}, A., {Bauer}, J., {Grav}, T., {Masiero}, J., {Cutri}, R.~M.,
  {Dailey}, J., et~al. (2011).
\newblock {Preliminary Results from NEOWISE: An Enhancement to the Wide-field
  Infrared Survey Explorer for Solar System Science}.
\newblock \emph{ApJ} 731, 53.
\newblock \doi{10.1088/0004-637X/731/1/53}
\bibAnnoteFile{Mainzer2011a}

\bibitem[{Mak{\'o} and Szenkovits(2004)}]{Mako2004}
Mak{\'o}, Z. and Szenkovits, F. (2004).
\newblock Capture in the circular and elliptic restricted three-body problem.
\newblock \emph{Celestial Mechanics and Dynamical Astronomy} 90, 51--58.
\newblock \doi{10.1007/s10569-004-5899-7}
\bibAnnoteFile{Mako2004}

\bibitem[{{Marzari} and {Scholl}(2013)}]{Marzari2013-EM-Trojans}
{Marzari}, F. and {Scholl}, H. (2013).
\newblock {Long term stability of Earth Trojans}.
\newblock \emph{Celestial Mechanics and Dynamical Astronomy} 117, 91--100.
\newblock \doi{10.1007/s10569-013-9478-7}
\bibAnnoteFile{Marzari2013-EM-Trojans}

\bibitem[{{Meech} et~al.(2017){Meech}, {Weryk}, {Micheli}, {Kleyna}, {Hainaut},
  {Jedicke} et~al.}]{Meech2017-Oumuamua-Nature}
{Meech}, K.~J., {Weryk}, R., {Micheli}, M., {Kleyna}, J.~T., {Hainaut}, O.~R.,
  {Jedicke}, R., et~al. (2017).
\newblock {A brief visit from a red and extremely elongated interstellar
  asteroid}.
\newblock \emph{\nat} 552, 378--381.
\newblock \doi{10.1038/nature25020}
\bibAnnoteFile{Meech2017-Oumuamua-Nature}

\bibitem[{{Micheli} et~al.(2012){Micheli}, {Tholen}, and
  {Elliott}}]{Micheli2012-2009BD}
{Micheli}, M., {Tholen}, D.~J., and {Elliott}, G.~T. (2012).
\newblock {Detection of radiation pressure acting on 2009 BD}.
\newblock \emph{\nat} 17, 446--452.
\newblock \doi{10.1016/j.newast.2011.11.008}
\bibAnnoteFile{Micheli2012-2009BD}

\bibitem[{{Micheli} et~al.(2013){Micheli}, {Tholen}, and
  {Elliott}}]{Micheli2013-2012LA}
{Micheli}, M., {Tholen}, D.~J., and {Elliott}, G.~T. (2013).
\newblock {2012 LA, an optimal astrometric target for radiation pressure
  detection}.
\newblock \emph{\icarus} 226, 251--255.
\newblock \doi{10.1016/j.icarus.2013.05.032}
\bibAnnoteFile{Micheli2013-2012LA}

\bibitem[{{Mommert} et~al.(2014){Mommert}, {Farnocchia}, {Hora}, {Chesley},
  {Trilling}, {Chodas} et~al.}]{Mommert2014}
{Mommert}, M., {Farnocchia}, D., {Hora}, J.~L., {Chesley}, S.~R., {Trilling},
  D.~E., {Chodas}, P.~W., et~al. (2014).
\newblock {Physical Properties of Near-Earth Asteroid 2011 MD}.
\newblock \emph{\apjl} 789, L22.
\newblock \doi{10.1088/2041-8205/789/1/L22}
\bibAnnoteFile{Mommert2014}

\bibitem[{{Morais} and {Morbidelli}(2002)}]{Morais2002-EarthCoOrbitals}
{Morais}, M.~H.~M. and {Morbidelli}, A. (2002).
\newblock {The Population of Near-Earth Asteroids in Coorbital Motion with the
  Earth}.
\newblock \emph{\icarus} 160, 1--9.
\newblock \doi{10.1006/icar.2002.6937}
\bibAnnoteFile{Morais2002-EarthCoOrbitals}

\bibitem[{{Nesvorn{\'y}} et~al.(2014){Nesvorn{\'y}}, {Vokrouhlick{\'y}}, and
  {Deienno}}]{Nesvorny2014-IrregularSatellites}
{Nesvorn{\'y}}, D., {Vokrouhlick{\'y}}, D., and {Deienno}, R. (2014).
\newblock {Capture of Irregular Satellites at Jupiter}.
\newblock \emph{\apj} 784, 22.
\newblock \doi{10.1088/0004-637X/784/1/22}
\bibAnnoteFile{Nesvorny2014-IrregularSatellites}

\bibitem[{{Nesvorn{\'y}} et~al.(2007){Nesvorn{\'y}}, {Vokrouhlick{\'y}}, and
  {Morbidelli}}]{Nesvorny2007-IrregularSatellites}
{Nesvorn{\'y}}, D., {Vokrouhlick{\'y}}, D., and {Morbidelli}, A. (2007).
\newblock {Capture of Irregular Satellites during Planetary Encounters}.
\newblock \emph{\aj} 133, 1962--1976.
\newblock \doi{10.1086/512850}
\bibAnnoteFile{Nesvorny2007-IrregularSatellites}

\bibitem[{{Ohtsuka} et~al.(2008){Ohtsuka}, {Ito}, {Yoshikawa}, {Asher}, and
  {Arakida}}]{Ohtsuka2008-147P}
{Ohtsuka}, K., {Ito}, T., {Yoshikawa}, M., {Asher}, D.~J., and {Arakida}, H.
  (2008).
\newblock {Quasi-Hilda comet 147P/Kushida-Muramatsu. Another long temporary
  satellite capture by Jupiter}.
\newblock \emph{\aap} 489, 1355--1362.
\newblock \doi{10.1051/0004-6361:200810321}
\bibAnnoteFile{Ohtsuka2008-147P}

\bibitem[{Perozzi et~al.(2017)Perozzi, Ceccaroni, Valsecchi, and
  Rossi}]{Perozzi2017-DRO}
Perozzi, E., Ceccaroni, M., Valsecchi, G.~B., and Rossi, A. (2017).
\newblock Distant retrograde orbits and the asteroid hazard.
\newblock \emph{The European Physical Journal Plus} 132, 367.
\newblock \doi{10.1140/epjp/i2017-11644-0}
\bibAnnoteFile{Perozzi2017-DRO}

\bibitem[{{Pollack} et~al.(1979){Pollack}, {Burns}, and {Tauber}}]{Pollack1979}
{Pollack}, J.~B., {Burns}, J.~A., and {Tauber}, M.~E. (1979).
\newblock {Gas drag in primordial circumplanetary envelopes - A mechanism for
  satellite capture}.
\newblock \emph{Icarus} 37, 587--611.
\newblock \doi{10.1016/0019-1035(79)90016-2}
\bibAnnoteFile{Pollack1979}

\bibitem[{Sanchez and McInnes(2011)}]{Sanchez2011}
Sanchez, J.~P. and McInnes, C.~R. (2011).
\newblock Asteroid resource map for near-earth space.
\newblock \emph{Journal of Spacecraft and Rockets} 48, 153--165.
\newblock \doi{10.2514/1.49851}
\bibAnnoteFile{Sanchez2011}

\bibitem[{Sanchez and McInnes(2013)}]{Sanchez2013}
Sanchez, J.~P. and McInnes, C.~R. (2013).
\newblock \emph{Asteroids: Prospective Energy and Material Resources.}
  (Springer, Berlin, Heidelberg), chap. 18 - Available Asteroid Resources in
  the Earth’s Neighbourhood.
\newblock 439--458.
\newblock \doi{10.1007/978-3-642-39244-3_18}
\bibAnnoteFile{Sanchez2013}

\bibitem[{{Schunov{\'a}-Lilly} et~al.(2017){Schunov{\'a}-Lilly}, {Jedicke},
  {Vere{\v s}}, {Denneau}, and {Wainscoat}}]{Schunova-Lilly2017}
{Schunov{\'a}-Lilly}, E., {Jedicke}, R., {Vere{\v s}}, P., {Denneau}, L., and
  {Wainscoat}, R.~J. (2017).
\newblock {The size-frequency distribution of H>13 NEOs and ARM target
  candidates detected by Pan-STARRS1}.
\newblock \emph{\icarus} 284, 114--125.
\newblock \doi{10.1016/j.icarus.2016.11.010}
\bibAnnoteFile{Schunova-Lilly2017}

\bibitem[{{Schwamb} et~al.(2018){Schwamb}, {Jones}, {Chesley}, {Fitzsimmons},
  {Fraser}, {Holman} et~al.}]{Schwamb2018-LSST-roadmap}
{Schwamb}, M.~E., {Jones}, R.~L., {Chesley}, S.~R., {Fitzsimmons}, A.,
  {Fraser}, W.~C., {Holman}, M.~J., et~al. (2018).
\newblock {Large Synoptic Survey Telescope Solar System Science Roadmap}.
\newblock \emph{ArXiv e-prints}
\bibAnnoteFile{Schwamb2018-LSST-roadmap}

\bibitem[{Semeniuk(2013)}]{Semeniuk2013}
[Dataset] Semeniuk, I. (2013).
\newblock Scientists yearn to catch a moon
\bibAnnoteFile{Semeniuk2013}

\bibitem[{{Sercel}(2017)}]{Sercel2017}
{Sercel}, J. (2017).
\newblock {Asteroid ISRU Concepts, Missions, and Technologies In Development}
\bibAnnoteFile{Sercel2017}

\bibitem[{{Shao} et~al.(2014){Shao}, {Nemati}, {Zhai}, {Turyshev}, {Sandhu},
  {Hallinan} et~al.}]{Shao2014-SyntheticTracking}
{Shao}, M., {Nemati}, B., {Zhai}, C., {Turyshev}, S.~G., {Sandhu}, J.,
  {Hallinan}, G., et~al. (2014).
\newblock {Finding Very Small Near-Earth Asteroids using Synthetic Tracking}.
\newblock \emph{\apj} 782, 1.
\newblock \doi{10.1088/0004-637X/782/1/1}
\bibAnnoteFile{Shao2014-SyntheticTracking}

\bibitem[{{Sidorenko} et~al.(2014){Sidorenko}, {Neishtadt}, {Artemyev}, and
  {Zelenyi}}]{Sidorenko2014-QuasiSatellites}
{Sidorenko}, V.~V., {Neishtadt}, A.~I., {Artemyev}, A.~V., and {Zelenyi}, L.~M.
  (2014).
\newblock {Quasi-satellite orbits in the general context of dynamics in the 1:1
  mean motion resonance: perturbative treatment}.
\newblock \emph{Celestial Mechanics and Dynamical Astronomy} 120, 131--162.
\newblock \doi{10.1007/s10569-014-9565-4}
\bibAnnoteFile{Sidorenko2014-QuasiSatellites}

\bibitem[{{Siltala} and {Granvik}(2017)}]{Siltala2017}
{Siltala}, L. and {Granvik}, M. (2017).
\newblock {Asteroid mass estimation using Markov-chain Monte Carlo}.
\newblock \emph{\icarus} 297, 149--159.
\newblock \doi{10.1016/j.icarus.2017.06.028}
\bibAnnoteFile{Siltala2017}

\bibitem[{{Stramacchia} et~al.(2016){Stramacchia}, {Colombo}, and
  {Bernelli-Zazzera}}]{Stramacchia2016-DRO}
{Stramacchia}, M., {Colombo}, C., and {Bernelli-Zazzera}, F. (2016).
\newblock {Distant Retrograde Orbits for space-based Near Earth Objects
  detection}.
\newblock \emph{Advances in Space Research} 58, 967--988.
\newblock \doi{10.1016/j.asr.2016.05.053}
\bibAnnoteFile{Stramacchia2016-DRO}

\bibitem[{{Strange} et~al.(2013){Strange}, {Landau}, {McElrath}, {Lantoine},
  {Lam}, {McGuire} et~al.}]{Strange2013-ARM}
{Strange}, N., {Landau}, D., {McElrath}, T., {Lantoine}, G., {Lam}, T.,
  {McGuire}, M., et~al. (2013).
\newblock {Overview of Mission Design for NASA Asteroid Redirect Robotic
  Mission Concept}.
\newblock In \emph{33rd International Electric Propulsion Conference
  (IEPC2013), Washington, D.C., October 6-10, 2013, id.4436}. 4436
\bibAnnoteFile{Strange2013-ARM}

\bibitem[{Szebehely(1967)}]{Szebehely1967}
Szebehely, V. (1967).
\newblock \emph{Theory of Orbits} (Academic Press, New-York.)
\bibAnnoteFile{Szebehely1967}

\bibitem[{Sánchez and Scheeres(2014)}]{Sanchez2014}
Sánchez, P. and Scheeres, D.~J. (2014).
\newblock The strength of regolith and rubble pile asteroids.
\newblock \emph{Meteoritics \& Planetary Science} 49, 788--811.
\newblock \doi{10.1111/maps.12293}
\bibAnnoteFile{Sanchez2014}

\bibitem[{{Takada}(2010)}]{Takada2010}
{Takada}, M. (2010).
\newblock {Subaru Hyper Suprime-Cam Project}.
\newblock In \emph{American Institute of Physics Conference Series}, eds.
  N.~{Kawai} and S.~{Nagataki}. vol. 1279 of \emph{American Institute of
  Physics Conference Series}, 120--127.
\newblock \doi{10.1063/1.3509247}
\bibAnnoteFile{Takada2010}

\bibitem[{{Tan} et~al.(2017){Tan}, {McInnes}, and
  {Ceriotti}}]{Tan2017-Direct+IndirectCapture}
{Tan}, M., {McInnes}, C., and {Ceriotti}, M. (2017).
\newblock {Direct and indirect capture of near-Earth asteroids in the
  Earth-Moon system}.
\newblock \emph{Celestial Mechanics and Dynamical Astronomy} 129, 57--88.
\newblock \doi{10.1007/s10569-017-9764-x}
\bibAnnoteFile{Tan2017-Direct+IndirectCapture}

\bibitem[{Tan et~al.(2017)Tan, McInnes, and
  Ceriotti}]{Tan2017-MomentumExchange}
Tan, M., McInnes, C.~R., and Ceriotti, M. (2017).
\newblock Low-energy near-earth asteroid capture using momentum exchange
  strategies.
\newblock \emph{Journal of Guidance, Control, and Dynamics} ,
  1--12\doi{10.2514/1.G002957}
\bibAnnoteFile{Tan2017-MomentumExchange}

\bibitem[{Tanikawa(1983)}]{Tanikawa1983}
Tanikawa, K. (1983).
\newblock Impossibility of the capture of retrograde satellites in the
  restricted three-body problem.
\newblock \emph{Celestial Mechanics and Dynamical Astronomy} 29, 367--402
\bibAnnoteFile{Tanikawa1983}

\bibitem[{Taylor et~al.(2018)Taylor, McDowell, and
  Elvis}]{Taylor2018-MainBeltDeltaV}
Taylor, A., McDowell, J.~C., and Elvis, M. (2018).
\newblock A delta-v map of the known main belt asteroids.
\newblock \emph{Acta Astronautica} 146, 73 -- 82.
\newblock \doi{https://doi.org/10.1016/j.actaastro.2018.02.014}
\bibAnnoteFile{Taylor2018-MainBeltDeltaV}

\bibitem[{{Taylor}(1995)}]{Taylor1995-MeteorVelocity}
{Taylor}, A.~D. (1995).
\newblock {The Harvard Radio Meteor Project Meteor Velocity Distribution
  Reappraised}.
\newblock \emph{\icarus} 116, 154--158.
\newblock \doi{10.1006/icar.1995.1117}
\bibAnnoteFile{Taylor1995-MeteorVelocity}

\bibitem[{Urrutxua and Bombardelli(2017)}]{Urrutxua2017}
Urrutxua, H. and Bombardelli, C. (2017).
\newblock A look at the capture mechanisms of the “temporarily captured
  asteroids” of the earth.
\newblock In \emph{26th International Symposium on Space Flight Dynamics}.
  ISSFD-2017-074, 1--7
\bibAnnoteFile{Urrutxua2017}

\bibitem[{Urrutxua et~al.(2015)Urrutxua, Scheeres, Bombardelli, Gonzalo, and
  Pel{\'a}ez}]{Urrutxua2015}
Urrutxua, H., Scheeres, D.~J., Bombardelli, C., Gonzalo, J.~L., and Pel{\'a}ez,
  J. (2015).
\newblock Temporarily captured asteroids as a pathway to affordable asteroid
  retrieval missions.
\newblock \emph{Journal of Guidance, Control, and Dynamics} 38, 2132--2145.
\newblock \doi{10.2514/1.G000885}
\bibAnnoteFile{Urrutxua2015}

\bibitem[{{Valdes} and {Freitas}(1983)}]{Valdes1983-EM-Trojans}
{Valdes}, F. and {Freitas}, R.~A. (1983).
\newblock {A search for objects near the earth-moon Lagrangian points}.
\newblock \emph{\icarus} 53, 453--457.
\newblock \doi{10.1016/0019-1035(83)90209-9}
\bibAnnoteFile{Valdes1983-EM-Trojans}

\bibitem[{Vieira-Neto and Winter(2001)}]{Vieira-Neto2001}
Vieira-Neto, E. and Winter, O.~C. (2001).
\newblock Time analysis for temporary gravitational capture: Satellites of
  uranus.
\newblock \emph{The Astronomical Journal} 122, 440
\bibAnnoteFile{Vieira-Neto2001}

\bibitem[{{Warren}(1994)}]{Warren1994-LunarMeteorites}
{Warren}, P.~H. (1994).
\newblock {Lunar and martian meteorite delivery services}.
\newblock \emph{\icarus} 111, 338--363.
\newblock \doi{10.1006/icar.1994.1149}
\bibAnnoteFile{Warren1994-LunarMeteorites}

\bibitem[{{Weber}(2005)}]{Weber2005-MeteorApparatus}
{Weber}, M. (2005).
\newblock {Some apparatuses of meteor astronomy from the pre-electronic epoch}.
\newblock \emph{WGN, Journal of the International Meteor Organization} 33,
  111--114
\bibAnnoteFile{Weber2005-MeteorApparatus}

\bibitem[{{Weryk} et~al.(2013){Weryk}, {Campbell-Brown}, {Wiegert}, {Brown},
  {Krzeminski}, and {Musci}}]{Weryk2013}
{Weryk}, R.~J., {Campbell-Brown}, M.~D., {Wiegert}, P.~A., {Brown}, P.~G.,
  {Krzeminski}, Z., and {Musci}, R. (2013).
\newblock {The Canadian Automated Meteor Observatory (CAMO): System overview}.
\newblock \emph{\icarus} 225, 614--622.
\newblock \doi{10.1016/j.icarus.2013.04.025}
\bibAnnoteFile{Weryk2013}

\end{thebibliography}


\section*{Figure captions}






\end{document}